# Cellular Mechanisms of Phase Maintenance in a Pyloric Motif of a Central Pattern Generator

**Gabrielle O'Brien[†1], Adam L. Weaver [†2], William H. Barnett[1], Dmitry A. Kozhanov[1], Gennady S. Cymbalyuk[1*]**
†: These authors made equal contributions.

[1]Neuroscience Institute, Georgia State University, Atlanta, Georgia, USA

[2]Departments of Biology and Neuroscience, St. Michael's College, Colchester, Vermont, USA

**\* Correspondence:**
Gennady S. Cymbalyuk
gcymbalyuk@gsu.edu

**Abstract**

In many neural networks, patterns controlling rhythmic behaviors are maintained across a wide range of periods. In the crustacean pyloric central pattern generator (CPG), a constant bursting pattern is preserved over a three-to-fivefold range of periods. We idescribe how neuromodulation could adjust neuronal properties to preserve phase relations as the period changes. We developed a biophysical model implementing a reduced pyloric network motif, which has a bursting neuron and two follower neurons interconnected through inhibitory synaptic coupling. We described cellular mechanisms supporting phase maintenance and investigated possible coordination between these mechanisms in four dynamically distinct ensembles of a pyloric CPG producing a triphasic pattern. The coordinated variation of the voltages of half-activation for potassium ($V_{K2}$) and hyperpolarization-activated ($V_h$) currents provides a family of three mechanisms for control of burst duration, interburst interval, and latency to spiking. The mechanisms are determined by the Cornerstone bifurcation, one of the Shilnikov blue sky catastrophe scenarios. In Mechanism 1, in a bursting neuron, the burst duration increases as $V_{K2}$ nears a blue-sky catastrophe bifurcation, while the interburst interval grows as $V_h$ approaches a saddle-node on an invariant circle bifurcation. In Mechanism 2, a silent neuron responds with a single burst to short input; the burst duration grows as $V_{K2}$ approaches a saddle-node bifurcation for periodic orbits. In Mechanism 3, a spiking neuron responds with a pause to short input; the pause duration grows as $V_h$ nears a saddle-node bifurcation for stationary states. In all three mechanisms, the measured quantities grow without bound as the bifurcation parameter nears its critical value, consistent with an inverse-square-root law.

## 1    Introduction

Much like the music of a well-rehearsed orchestra, biological rhythms such as leech swimming [1], lamprey swimming [2, 3], beating rhythm of the crayfish swimmerets [4, 5], ventilation in crabs [6] and stomach filtering and maceration in crustaceans [7, 8] remain coordinated over a wide range of cycle periods. An orchestra requires a reliable, rhythmically inclined conductor and musicians who play the right notes at the right time to maintain coherence. For neuronal networks, the mechanisms are less obvious: how can the CPG network maintain the precise functional "identity" of the activity pattern even as its speed changes? This property requires phase (i.e., burst duration / cycle period) maintenance: that each note in a melody, or activity of a neuron in a firing pattern, changes its temporal characteristics relative to the other pattern elements so that its relative timing (phase) is preserved. In CPGs, it demands that the duration of each neuron's activity and delay to firing after stimulation is regulated relative to the activity of the other neurons in the circuit. In many systems, the pace is regulated over a wide cycle period range as biological needs warrant by a driver neuron [9] or a timing network [10-14] with a set rhythmic activity. In the pyloric network, the follower neurons produce bursting activity within the interburst interval of the pacemaker to complete the functional pattern. Neuronal properties are not uniform within the pyloric network. While the driver neuron is endogenously bursting, across preparations the isolated follower neurons exhibit a variety of endogenous regimes of activity: silent, bursting, and spiking. Interestingly, many follower neurons burst rhythmically under certain modulatory conditions and hence are termed conditional bursters [15, 16]. The cellular mechanisms and dynamics that allow conditional bursters with apparently different dynamics to participate in phase-maintaining patterns are still a major area of investigation [17-21].

Previous studies of central pattern generators identified a number of synaptic [22-26] and cellular mechanisms [27, 28] for phase maintenance. There is a wealth of evidence for the essential role of cellular mechanisms in promoting functional phasing. The relative timing of neuronal activity across cycle periods is largely held constant due to compensatory changes in certain ion channel expression [29, 30], although there may be variation in network period and phase relationships across individuals with the same circuitry [31-33], and follower neurons may vary their latency to firing with the duration of driving input even when isolated from a network [34]. The ability of follower neurons to adjust their delay to firing is crucial to phase maintenance: in the absence of such flexibility, the phase between driver and follower will shrink as period grows. Control of follower neuron's temporal characteristics can be accomplished with slow gating variables of certain ionic currents, such as the A-type potassium current and the hyperpolarization-activated (h-) current, which have been shown to underlie both delay to excitation and burst duration in CPG neurons [27, 35]. The neuromodulatory inputs to the pyloric CPG can coregulate multiple ion currents (e.g., A- and h-type) to ensure functional motor patterns, suggesting a central role for intrinsic currents in phase maintenance [36]. Also, the coordination of various intrinsic currents can promote phase maintenance via interspike interval delays within bursts [28]. Neuronal activity patterns have been shown to regulate correlated ion channel mRNA levels [37]. Recent work by O'Leary et al. describes mechanisms explaining how neurons could regulate ion channel expression to produce an intrinsic "activity set point." Their model argues that both synaptic and intrinsic conductances can thus be regulated and promote network homeostasis [38, 39]. Taken together, these data suggest an important role for coordinated regulation of ionic currents in the phase-maintenance of neuronal patterns.

Here, we build on the theory of coregulation of ion currents as a means of maintaining the overall dynamical identity of a neuron with slow conductances that regulate opposite phases of the bursting cycle: silence and spiking. We suggest that minimally a family of mechanisms controlling two intrinsic currents may explain experimental results from the standpoint of cellular dynamics. We also consider how the coordination of these currents allows conditional bursters to maintain phase with an endogenously bursting driver, a question relevant to a variety of CPG systems [17]. To present a convincing case in a biologically relevant setting, we turn to the crustacean pyloric network to investigate the feasibility of intrinsic current modulation to contribute to phase maintenance in a pyloric CPG motif.

Central pattern generators of invertebrates are exceptional targets for phase maintenance studies because they are typically small networks with neurons that are identifiable across specimens [15, 19, 40-44]. These qualities make them ideal for both *in vitro* studies and computational modeling. The pyloric network is of particular note: it is a sub-network in the larger stomatogastric system and consists of a small, well-studied circuit that controls filtration and supports grinding of food particles in crustaceans and is known to exhibit phase maintenance with three-to-five fold variation in period [7, 8, 24, 34, 45, 46]. This circuit has been shown to consist of an endogenous bursting pacemaker and conditional oscillator follower neurons that together produce a triphasic rhythm. These follower neurons typically require central neuromodulatory inputs and regular synaptic inhibition to produce regular bursts of activity [15, 16]. We consider a reduced model CPG whose circuitry is derived from the pyloric network [24]. In the model, an endogenously bursting pacemaker (AB) neuron inhibits two follower (LP & PY) neurons, which are in turn reciprocally inhibiting each other. The structure of this circuit is conducive to a multiphasic pattern and may be a subset of a larger rhythmic network. We refer to it as a motif, a configuration of neurons and synapses that may appear in a variety of biological systems and is not species or function-specific [47].

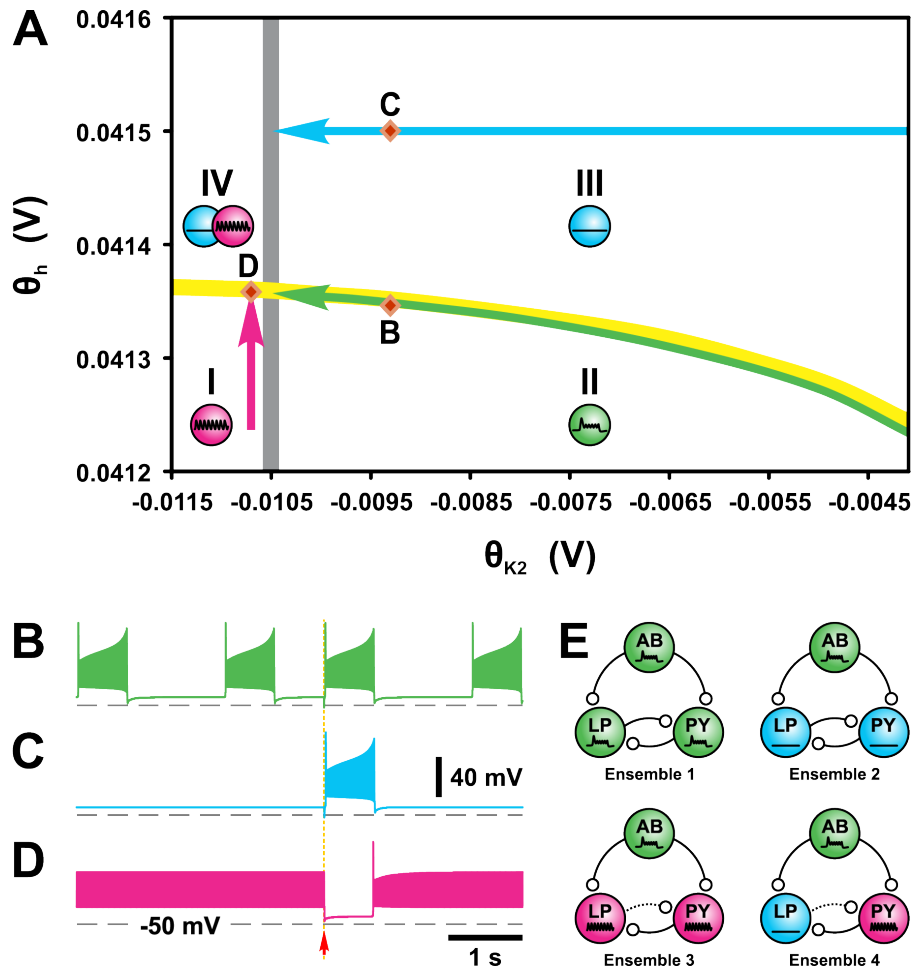

**Fig. 1 The global codimension-2 Cornerstone bifurcation provides a family of mechanisms for control of the temporal characteristics of endogenous bursting and pulse triggered responses. (A)** Regimes of single neuron activity are mapped on the plane of varied model parameters $\theta_{K2}$ and $\theta_h$. Two bifurcation curves partition the plane into four regions of distinct activity. The grey curve between regions I and II is a blue-sky catastrophe. The grey curve between regions III and IV is a saddle-node bifurcation for periodic orbits. The yellow curve between regions II and III is a SNIC bifurcation. The intersection of the grey and black curves is the Cornerstone bifurcation. The yellow curve between regions I and IV is a saddle-node bifurcation for equilibria. Region I corresponds to a spiking regime, region II corresponds to a bursting regime, region III corresponds to a silent regime and region IV corresponds to a bistable regime with coexistence of spiking and silence. The green arrow traces the used parameters along an endogenous bursting regime arc just below the yellow bifurcation curve with a cycle period ranging from 0.5 (base) to 30 (tip) s. The blue arrow is the matched parameter trajectory for the silent regime and the red arrow matches with tonic firing parameters. [modified from Fig. 1A in Barnett and Cymbalyuk [48]] **(B-D)** Each isolated neuron model was integrated for 100 s without current injection to reach the attractor, then a brief (1 ms) hyperpolarizing pulse (0.2 nA in amplitude) was applied to perturb the cell and transient response without any injected current was recorded for 33.3 s. **(B)** A bursting neuron with parameters close to the SNIC bifurcation ($\theta_h$ = 0.04134595 V, $\theta_{K2}$ = -0.0093 V) has a cycle period of 1.985 s, burst duration of 0.651 s, interburst interval of 1.333 s, and duty cycle of 0.328. **(C)** A silent neuron with parameters above the SNIC bifurcation ($\theta_h$ = 0.0415 V, $\theta_{K2}$ = -0.0093 V) was induced to produce a single burst of 0.650 s. **(D)** A tonic spiking neuron with parameters below the saddle-node bifurcation for equilibria ($\theta_h$ = 0.041356765583 V, $\theta_{K2}$ = -0.0106999 V) was induced to produce a latency to spiking of 0.656 s. **(E)** The pyloric circuitry is configured as four distinct ensembles distinguished by the activity regime of the isolated follower neurons. In all Ensembles, the driver neuron AB is configured as endogenously bursting. In Ensemble 1, the follower neurons are also endogenously bursting. In Ensemble 2, the follower neurons are endogenously silent. In Ensemble 3, the follower neurons are endogenously spiking. In Ensemble 4, the follower neuron LP is endogenously silent and the follower neuron PY is endogenously spiking. The dashed synapse from LP to PY indicates that robust triphasic rhythm is established if the strength of LP to PY is very weak or zero.

We suggest that bifurcations underlying the transitions between the three basic regimes of neuronal activity (silence, bursting, and spiking) could provide complementary basic rules organizing phase maintenance of the functional pattern around a co-dimension 2 bifurcation as an organizing center. This bifurcation is an example of Shilnikov blue-sky catastrophe scenarios [49-51]. We have previously described a family of intrinsic cellular mechanisms that explain the coregulation of intrinsic cellular currents in terms of the theory of dynamical systems. Through the coregulation of a potassium ($I_{K2}$) and hyperpolarization-activated ($I_h$) current, we achieve control of bursting activity and pulse-triggered responses in individual neurons [48]. We focus on two parameters: $\theta_{K2}$ and $\theta_h$ which describe the voltages of half-activation of $I_{K2}$ and $I_h$ (Fig. 1A). Under smooth variation of the voltage of half-activation of $I_{K2}$, our model neuron undergoes a blue-sky catastrophe bifurcation, which controls its transition from bursting to spiking. Independently, variation of the voltage of half-activation of $I_h$ leads to a saddle-node bifurcation on invariant circle (SNIC) bifurcation, which controls the transition from bursting to silence. When the criteria for both bifurcations are satisfied, a global codimension-2 bifurcation occurs [48-51]. The choices of the two parameters in regions around this critical point determine a neuron's endogenous activity: bursting, silent, or spiking. As the transition between regions is approached, the corresponding temporal characteristic of a neuron changes in a precisely controllable manner: it grows proportionally to $\frac{1}{\sqrt{|\alpha - \alpha^*|}}$ where $\alpha^*$ is the bifurcation value of $\alpha$ and $\alpha > \alpha^*$. We called this co-dimension 2 bifurcation the Cornerstone bifurcation [48], since by precise navigation of the two bifurcation parameters in its vicinity, the burst duration and interburst interval of an endogenously bursting neuron can be independently set from very small to nearly any arbitrarily large values. Similarly, pulse-triggered responses in endogenously bursting, silent, and spiking neurons can be controlled. When a short current pulse stimulus is applied to a bursting neuron, a phase-advanced burst of the same duration can be induced (Fig. 1B). Similarly, a pulse of current applied to a silent neuron close to the border of silence and bursting triggers a single stereotyped burst whose duration is set by the Cornerstone bifurcation (Fig. 1C). Finally, a current pulse applied to a spiking neuron close to the border of spiking and bursting silences the neuron for a determined interval of time whose duration is controlled analogously (Fig. 1D). This provides three potential mechanisms for control of temporal characteristics in neurons, one for each endogenous activity regime.

To demonstrate the applicability of these mechanisms to phase maintenance in the driver-follower motif, we apply them independently and in conjunction to orchestrate triphasic phase maintaining patterns in four ensembles of neurons (Fig. 1E). The ensembles all include an endogenously bursting driver neuron, but differ in the selection of paired endogenously bursting, silent, or spiking follower neurons. The combination of ensembles demonstrates the wide applicability of the family of mechanisms to biological systems where the endogenous activities of follower neurons are heterogeneous or lack pacemaker qualities. We show how the mechanisms support phase maintenance in ensembles that are homogenous bursters (Ensemble 1) as well as ensembles with conditionally bursting follower neurons, which are here either silent (Ensemble 2) or spiking endogenously (Ensemble 3). We also evaluate a fully heterogeneous network motif (Ensemble 4) with an endogenous bursting pacemaker (AB) neuron, one silent follower (LP) neuron, and one tonic firing follower (PY) neuron. We chose this configuration as it is closest to the activity patterns found in the pyloric network in the absence of central neuromodulatory inputs [16]. These conditional bursters are of interest for their ability to participate in a flexible phase maintaining neural circuit although their intrinsic dynamics do not normally support rhythmic activity. The impact of the mechanism on the model's behavior is observed through the changes of average $I_h$ and $I_{K2}$ in the interburst and burst intervals, respectively, across 20 cycle period trajectories (0.5 to 30 s). Thus, we provide evidence that our general neuronal model can be readily tuned for specific functional networks.

## 2    Materials and Methods

### 2.1    Model

Previously, we developed a Hodgkin-Huxley style neuronal model with three voltage gated currents: a fast $Na^+$ current, $I_{Na}$, a non-inactivating $K^+$ current, $I_{K2}$, and a hyperpolarization-activated current, $I_h$ [48]. This model was inspired by electrophysiological experiments with the leech heart interneuron under physiological conditions when the synaptic currents, the $Ca^{2+}$ currents, and most of the $K^+$ currents are blocked with $Co^{2+}$ and 4-Aminopyridine and the leech heart interneuron exhibits plateau-like bursts oscillating on a time scale of tens of seconds [52, 53]. To bring the time scale of the model into agreement with the time scale of pyloric neurons, we matched the width of spikes by scaling the model time. In the original model [48], average width of spikes was ~ 209 ms, while experimental recordings of spike width in the AB, LP and VD neurons of the pyloric network in the living system showed ~ 7 ms [54, 55]. To match the model spike widths, a time scaling factor $\chi$ ($\chi$=30) was introduced into all model equations to adjust temporal characteristics. Thus, the cycle period of the original model range of 15 to 900 s scaled down to a more biologically relevant range of 0.5 to 30 s.

Each cell $i$, $i\epsilon\{AB, LP, PY\}$, obeys the following equations describing the dynamics of our pyloric motif:

$$C\frac{dV_i}{dt} = -\chi[\bar{g}_{Na}m_{Na,\infty,i}(V_i)^3 h_{Na,i}[V_i - E_{Na}] + \bar{g}_{K2}m_{K2,i}{}^2[V_i - E_K] + \bar{g}_h m_{h,i}{}^2[V_i - E_h]$$
$$+ g_{leak}[V_i - E_{leak}] + 0.006 + I_{synTotal,i} - I_{Inject}]$$

$$\frac{dh_{Na,i}}{dt} = \chi\left[\frac{1}{1 + \exp(500[V_i + 0.0325])} - h_{Na,i}\right]/0.0405$$

$$\frac{dm_{h,i}}{dt} = \chi\left[\frac{1}{1 + 2\exp(180[V_i + \theta_h]) + \exp(500[V_i + \theta_h])} - m_{h,i}\right]/0.1$$

$$\frac{dm_{K2,i}}{dt} = \chi\left[\frac{1}{1 + \exp(-83[V_i + \theta_{K2}])} - m_{K2,i}\right]/2$$

where:

$$m_{Na,\infty}(V) = \frac{1}{1 + \exp(-150[V + 0.0305])}$$

Synapses are appended to the voltage differential equation as follows:

$$I_{syn,j\rightarrow i} = \bar{g}_{j\rightarrow i}s_j(V_i - E_{syn})$$

Where $s_j$ is synaptic activation variable of the presynaptic cell $j$, $j \epsilon \{AB, LP, PY\}$ and is governed by:

$$\frac{ds_j}{dt} = \chi \left[ \frac{1}{1 + \exp(-5000[V_j + 0.02])} - s_j \right] / 0.015$$

In our motif, the pacemaker cell AB does not receive any synaptic input:

$$I_{synTotal,AB} = 0 \; nA$$

and each follower cell receives input from AB and the other follower cell:

$$I_{synTotal,LP} = I_{syn,AB \to \text{LP}} + I_{syn,PY \to \text{LP}}$$

$$I_{synTotal,PY} = I_{syn,AB \to PY} + I_{syn,\text{LP} \to PY}$$

The membrane potential of the neuron i is denoted by $V_i$. The activation of $I_{Na}$ is instantaneous and is denoted as $m_{Na,\infty}$. The inactivation of $I_{Na}$ and the activations of $I_{K2}$ and $I_h$ are $h_{Na}$, $m_{K2}$ and $m_h$. For consistency with our previous publications [48, 49], we use the regulating parameters $\theta_{K2}$ and $\theta_h$ corresponding to the voltages of half-activation of the variables $m_{K2}$ and $m_h$ ($V_{\frac{1}{2}K2}$ = -$\theta_{K2}$ and $V_{\frac{1}{2}h}$ = -$\theta_h$). The function $m_{K2,\infty}(V)$ is the steady state activation function of $I_{K2}$. The slow variable of this system is $m_{K2}$. The synaptic activation function is represented by $s$ where $V_j$ is the membrane potential of the pre-synaptic neuron j.

Cellular parameters are given in Table 1. Synaptic parameters are provided in Table 2.

**Table 1 Parameters of the model neuron.**

| Symbol | Description | Value |
|---|---|---|
| C | Membrane capacitance | 2 nF |
| $\bar{g}_{Na}$ | Conductance of sodium current | 105 nS |
| $\bar{g}_{K2}$ | Conductance of potassium current | 30 nS |
| $\bar{g}_h$ | Conductance of hyperpolarization-activated current | 4 nS |
| $g_{leak}$ | Conductance of leak current | 8 nS |
| $E_{Na}$ | Reversal potential of sodium current | 0.045 V |
| $E_K$ | Reversal potential of potassium current | -0.070 V |
| $E_h$ | Reversal potential of hyperpolarization-activated current | -0.021 V |
| $E_{leak}$ | Reversal potential of leak current | -0.046 V |
| $\theta_h$ | Negative value of the voltage of half-activation of the hyperpolarization-activated current | Bifurcation parameter |
| $\theta_{K2}$ | Negative value of the voltage of half-activation of the potassium current | Bifurcation parameter |

**Table 2 Parameters of synaptic currents**

| Symbol | Description | Ensembles 1 & 2 | Ensembles 3 & 4 |
|---|---|---|---|
| $\bar{g}_{AB \to LP}$ | Conductance of AB→LP synapse | 50 nS | 50 nS |
| $\bar{g}_{AB \to PY}$ | Conductance of AB→PY synapse | 10 nS | 10 nS |
| $\bar{g}_{LP \to PY}$ | Conductance of LP→PY synapse | 50 nS | 0 nS |
| $\bar{g}_{PY \to LP}$ | Conductance of PY→LP synapse | 1 nS | 50 nS |
| $E_{AB \to LP}$ | Reversal pot. of AB→LP synaptic current | -0.048 V | Ensemble 3: -0.04214 V<br>Ensemble 4: -0.048 V |
| $E_{AB \to PY}$ | Reversal pot. of AB→PY synaptic current | -0.048 V | -0.04214 V |
| $E_{LP \to PY}$ | Reversal pot. of LP→PY synaptic current | -0.048 V | N/A |
| $E_{PY \to LP}$ | Reversal pot. of PY→LP synaptic current | -0.048 V | -0.048 V |

The activity regimes of an isolated neuron are determined by $\theta_{K2}$ and $\theta_h$. The two-parameter bifurcation diagram is partitioned into four distinct regions with distinct activity (Fig. 1A). For parameters in Region I (low values of $\theta_{K2}$ and $\theta_h$), a neuron is endogenously spiking. In Region II (high values of $\theta_{K2}$ and low values of $\theta_h$), a neuron is endogenously bursting. In Region III (high values of $\theta_{K2}$ and $\theta_h$), a neuron is endogenously silent. In Region IV (low values of $\theta_{K2}$ and high values of $\theta_h$), a neuron exhibits bistability of endogenous silence and spiking. In the bistable region, the activity regime of the neuron is dependent on initial conditions.

Regions I and II are separated by a blue-sky catastrophe which marks the transition from bursting to spiking (occurs at approximately $\theta_{K2} = -0.0105$ V, almost regardless of $\theta_h$). Similarly, Regions III and IV are separated by a saddle-node bifurcation for periodic orbits. Regions II and III are separated by a SNIC bifurcation, which marks the transition from bursting to silence (occurs at approximately $\theta_h = 0.0413$ V and is weakly dependent on $\theta_{K2}$). Similarly, Regions I and IV are separated by a saddle-node bifurcation for equilibria. The Cornerstone bifurcation occurs when the criteria for both bifurcations are satisfied (approximately $\theta_{K2}{}^* = -0.010505$ V, $\theta_h{}^* = 0.041356548$ V). These values were determined in our previous work [48] by sampling the activity on a grid in steps of 0.00004 V.

## 2.2 Mechanisms for Precise Temporal Control of Endogenous and Pulse-Triggered Activity

Based on the results of our model, we describe three proposed mechanisms for precise temporal control of endogenous and pulse-triggered activity of endogenously spiking, bursting, and silent neurons, found in the monostable regions I, II, and III, respectively (Fig. 1A).

**Mechanism 1 controls the burst duration and interburst interval of a bursting neuron.** As the potassium current becomes less and less activated during the burst phase of an endogenously bursting neuron, termination of the burst is postponed. In our model, as $\theta_{K2}$ of a bursting neuron approaches a critical value for bifurcation, the burst duration grows arbitrarily large according to the inverse-square-root law. When $\theta_{K2}$ crosses the critical value, the neuron becomes tonically spiking at the blue-sky catastrophe [49]. Similarly, when $\theta_h$ approaches its critical value for the SNIC bifurcation, the interburst interval of a bursting neuron grows arbitrarily large according to the inverse-square-root law [48]. When $\theta_h$ crosses the critical value, the neuron becomes silent in a SNIC bifurcation [56]. When both parameters are at their critical value, the two bifurcation curves intersect at a global codimension-2 bifurcation, the Cornerstone bifurcation [48]. The duty cycle (i.e., burst duration / cycle period) of a bursting neuron is determined by the relative positions of $\theta_{K2}$ and $\theta_h$ (i.e., analogous to coregulation) compared with the Cornerstone bifurcation (Fig. 1A, green arrow).

**Mechanism 2 controls the duration of the single burst produced by a silent neuron in response to synaptic inhibitory input.** Near the saddle-node bifurcation for periodic orbits, a silent neuron demonstrates transient spiking activity at the bifurcation parameter values where two periodic orbits, stable and unstable, are about to appear. This spiking activity is a "ghost" of the spiking about to be born and forms a single burst, the duration of which is determined by the difference of the parameter value and its bifurcation value. The mechanism that controls the burst duration of a silent neuron is similar to the control of burst duration in the bursting neuron: varying $\theta_{K2}$ toward the saddle-node bifurcation for periodic orbits makes the potassium current less activated and burst duration grows arbitrarily large (Fig. 1A, blue arrow).

**Mechanism 3 controls the duration of the pause in spiking activity of a tonically spiking neuron in response to synaptic inhibitory input.** Near the saddle-node bifurcation for equilibria, a tonically spiking neuron demonstrates transient silent activity at the bifurcation parameter values where two equilibria, stable and unstable, are about to appear. This pause in spiking activity is a "ghost" of the rest state about to be born and forms a silent interval, the duration of which is determined by the difference of the parameter value and its bifurcation value. The mechanism controlling the duration of the pause to spiking is parallel to the control of interburst interval in the bursting neuron: varying $\theta_h$ toward a saddle-node bifurcation for equilibria makes the pause grow arbitrarily long (Fig. 1A, pink arrow).

Here we present a generic model of the pyloric motif. It contains three neurons, AB, LP and PY (shorthand for the anterior burster, lateral pyloric, and pyloric neurons) connected by four inhibitory synapses: AB→LP, AB→PY, LP→PY, and PY→LP. The pyloric network exhibits a triphasic burst pattern whose period is set by the endogenously bursting pacemaker AB neuron. As such, each AB burst is followed by an LP burst and then a PY burst. Each neuron is represented by our model and the inhibitory synapses are described by a conductance-based equation with synaptic activation modeled by a sigmoid function. We examined the dynamics of a family of four ensembles having the same pyloric synaptic connectivity motif but different endogenous properties of LP and PY with AB always endogenously bursting (Fig. 1E). In Ensemble 1, an endogenously bursting driver serves as the pacemaker for two endogenously bursting followers. In Ensemble 2, an endogenously bursting driver serves as the pacemaker for two endogenously silent followers. In Ensemble 3, an endogenously bursting driver serves as the pacemaker for two endogenously spiking followers. In Ensemble 4, an endogenously bursting driver serves as the pacemaker for one endogenously silent follower (LP) neuron and one endogenously spiking (PY) follower neuron. Our method for generating triphasic patterns is as follows: first, we used Mechanism 1 to give AB approximately one-third duty cycle at each cycle period. Second, we coordinated LP and PY using Mechanism 1, 2 or 3 to give both followers approximately one-third duty cycle. Third, we selected synaptic weighting that produced the correct follower neuron burst order. This method was repeated selecting the AB neuron to have approximately one-third duty cycle at a new cycle period, and then coordinating the follower neurons as in the second step while attempting to limit how many parameters were varied in the followers while still achieving a triphasic rhythm.

The system of equations (1) was integrated with the explicit embedded Runge-Kutta Prince-Dormand (8,9) method in the GSL GNU library (error tolerances: absolute $10^{-9}$, relative $10^{-9}$) or in MATLAB with the stiff ode15s solver (error tolerances: absolute $10^{-9}$, relative $10^{-8}$). In each simulation, we integrated the system of ODEs for at least 1000 seconds before collecting data to allow the model to stabilize near its attractor. For Fig. 1 B-D, an isolated neuron model was integrated piecewise for 100 s without current injection to let the neuron reach the attractor, then integrated for a brief (1 ms) period with a hyperpolarizing pulse (0.2 nA in amplitude) to "kick" the model away from the attractor, and finally integrated without any injected current for an additional 33.3 s. In each integration case for Fig. 1 B-D, the final parameters of each piecewise integration served as the initial conditions for the next round of integration.

## 3    Results

As a major motivation of this study is to determine the constraints that follower neuron intrinsic properties place on functional pattern generation in a model of the pyloric motif, we explored a variety of network configurations with different follower neuron firing properties (i.e., bursting, silent, spiking).

### 3.1    Phase-maintenance in Ensemble 1 where the two follower neurons are endogenously bursting

In Ensemble 1, the driver and both its followers are endogenously bursting. The desired pattern is achieved when each neuron spikes for approximately a third of the total period. The follower neurons do not require any input to produce activity, but they must fire in the correct order and with bursts of the appropriate duration. Mechanism 1 enables bursting with any required duty cycle to be realized at effectively any period. Through coordination of both $\theta_{K2}$ and $\theta_h$ between the SNIC bifurcation and blue-sky catastrophe, a duty cycle of approximately one-third was set for each neuron in the circuit at a range of periods (Fig. 1A, green arrow). We first used Mechanism 1 to select parameters for the driver AB that yielded endogenous bursting with approximately one-third duty cycle and period 0.5 s ($\theta_h$=0.04123 V, $\theta_{K2}$ = -0.0041 V). We then applied Mechanism 1 to the follower neurons so that they would also endogenously burst with one-third duty cycle and a cycle period of 0.5 s (Fig. 2 A1).

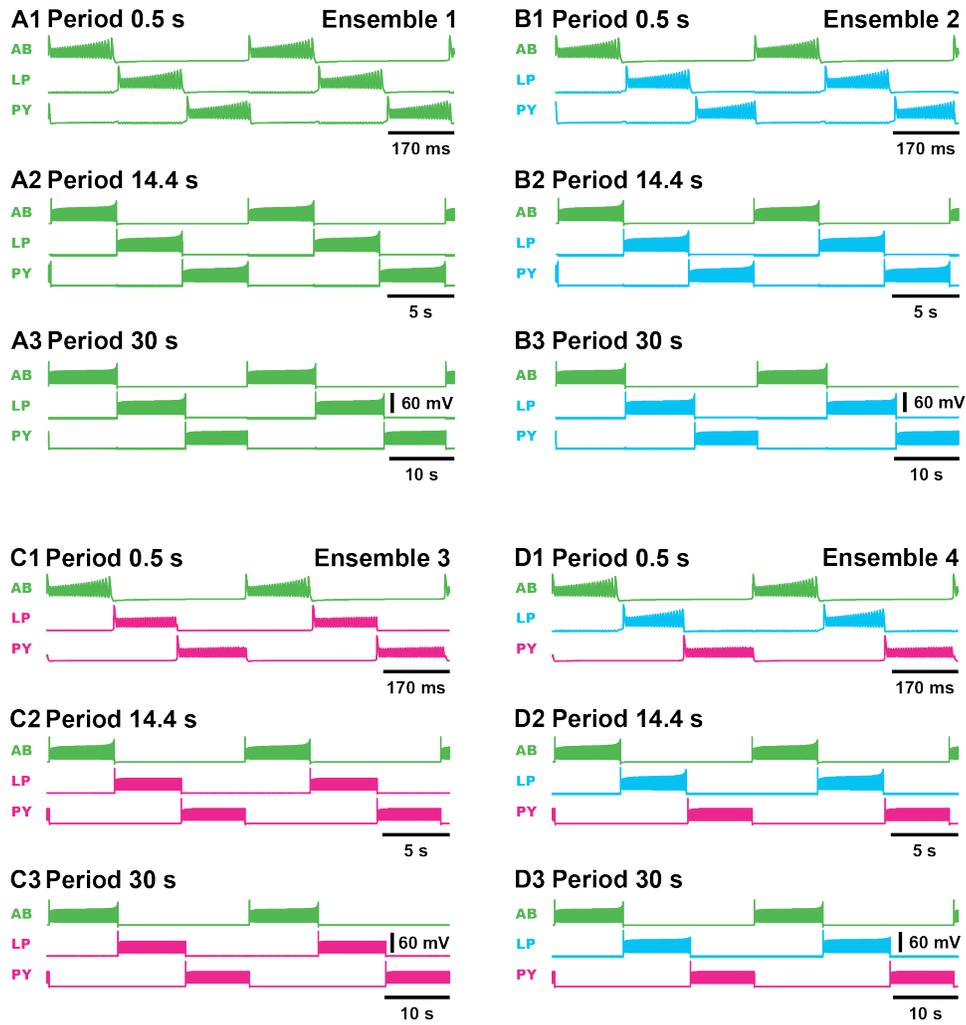

**Fig. 2 Three representative patterns for the four network ensembles.** The patterns are shown at periods of 0.5, 14.4, and 30 s. **(A) Endogenously bursting follower neurons (Ensemble 1).** Coordinates for all neurons in **(A1)** are ($\theta_h$ = 0.04123 V, $\theta_{K2}$ = -0.0041 V), coordinates for all neurons in **(A2)** are ($\theta_h$ = 0.04135621058 V, $\theta_{K2}$ = -0.0104655 V), and coordinates for all neurons in **(A3)** are ($\theta_h$ = 0.04135646145 V, $\theta_{K2}$ = -0.010496 V). **(B) Endogenously silent follower neurons (Ensemble 2).** Coordinates for **(B1)** are (AB: $\theta_h$ = 0.04123 V, $\theta_{K2}$ = -0.0041 V; LP: $\theta_h$ = 0.0415 V, $\theta_{K2}$ = -0.0041 V; PY: $\theta_h$ = 0.0415 V, $\theta_{K2}$ = -0.0041 V). Coordinates for **(B2)** are (AB: $\theta_h$ = 0.04135621058 V, $\theta_{K2}$ = -0.0104655 V; LP: $\theta_h$ = 0.040 V, $\theta_{K2}$ = -0.0104655 V; PY: $\theta_h$ = 0.0415 V, $\theta_{K2}$ = -0.0104655 V). Coordinates for **(B3)** are (AB: $\theta_h$ = 0.04135646145 V, $\theta_{K2}$ = -0.010496 V; LP: $\theta_h$ = 0.0415 V, $\theta_{K2}$ = -0.010496 V; PY: $\theta_h$ = 0.0415 V, $\theta_{K2}$ = -0.010496 V). **(C) Endogenously spiking follower neurons (Ensemble 3).** Coordinates for **(C1)** are (AB: $\theta_h$ = 0.04123 V, $\theta_{K2}$ = -0.0041 V; LP: $\theta_h$ = 0.040 V, $\theta_{K2}$ = -0.0107 V; PY: $\theta_h$ = 0.04134 V, $\theta_{K2}$ = -0.0106999 V). Coordinates for **(C2)** are (AB: $\theta_h$ = 0.04135621058 V, $\theta_{K2}$ = -0.0104655 V; LP: $\theta_h$ = 0.040 V, $\theta_{K2}$ = -0.0107 V; PY: $\theta_h$ = 0.041358044 V, $\theta_{K2}$ = -0.010699 V). Coordinates for **(C3)** are (AB: $\theta_h$ = 0.04135646145; V, $\theta_{K2}$ = -0.010496 V; LP: $\theta_h$ = 0.040 V, $\theta_{K2}$ = -0.0107 V; PY: $\theta_h$ = 0.04135804605 V, $\theta_{K2}$ = -0.0106999 V). **(D) Endogenously silent follower LP neuron and endogenously spiking follower PY neuron (Ensemble 4).** Coordinates for **(D1)** are (AB: $\theta_h$ = 0.04123 V, $\theta_{K2}$ = -0.0041 V; LP: $\theta_h$ = 0.0415 V, $\theta_{K2}$ = -0.0041 V; PY: $\theta_h$ = 0.04134 V, $\theta_{K2}$ = -0.0106999 V). Coordinates for **(D2)** are (AB: $\theta_h$ = 0.04135621058 V, $\theta_{K2}$ = -0.0104655 V; LP: $\theta_h$ = 0.0415 V, $\theta_{K2}$ = -0.0107 V; PY: $\theta_h$ = 0.041358044 V, $\theta_{K2}$ = -0.010699 V). Coordinates for **(D3)** are (AB: $\theta_h$ = 0.04135646145 V, $\theta_{K2}$ = -0.010496 V; LP: $\theta_h$ = 0.0415 V, $\theta_{K2}$ = -0.0107 V; PY: $\theta_h$ = 0.04135804605 V, $\theta_{K2}$ = -0.0106999 V).

Next, we established the ensemble model with the proper bursting order of the circuit pattern. We roughly adjusted the synapses primarily to break symmetry of the network and ensure that the follower neurons would not fire synchronously within the interval bounded by the pacemaker AB bursts. We set a stronger inhibitory synaptic coupling from AB to LP than from AB to PY (50 nS versus 10 nS). Additionally, the strength of the reciprocal inhibition between the follower neurons was asymmetrically weighted to ensure that PY remained hyperpolarized for the entirety of LP's burst. It was not necessary to have strong inhibition from PY to LP, as LP's burst had a fixed duration and did not encroach on PY's activity when duty cycle was set appropriately. We subsequently reapplied Mechanism 1 to AB, moving $\theta_{K2}$ and $\theta_h$ towards the Cornerstone bifurcation in a concave arc such that endogenous bursting with approximately one-third duty cycle occurred at longer periods (green arrow in Fig. 1A). We determined 20 sets of parameters that satisfied this criterion (supplementary Table 1). The parameter sets produce bursting with periods ranging from 0.5 to 30 s (examples shown in Fig. 2 A1-A3); in the collection, periods are at most 2.13 seconds apart. We then recorded a triphasic pattern at each period by fixing identical parameters in the follower neurons to that of the driver neuron (Fig. 3). By design, driver neuron burst duration scaled linearly with period (Fig. 3A). If the interburst interval and burst duration were set to exactly a 2:1 ratio, the duty cycle of the driver neuron would be kept exactly 0.333. We computed a linear regression for the burst duration of the driver against pattern period with a slope of 0.3361. This indicates that for all parameter sets, which supported a wide range of cycle period, the burst duration of the endogenously bursting neurons was approximately a third of the total period. Interburst interval also scaled linearly with period as determined by Mechanism 1 (Fig. 3A), and we computed a linear regression for the interburst interval of the driver with slope of 0.6639. This indicates that for all parameter sets, the interburst interval of the endogenously bursting driver neuron was approximately two thirds of the total period. The number of spikes per burst scaled linearly with period (Fig. 3B), indicating that the mechanism for increasing burst duration functions through the addition of spikes to each burst. This result is in accordance with the blue-sky catastrophe properties, which predict such addition of spikes with roughly the same waveform and thus with the same interspike interval [48]. We computed a linear regression for the number of spikes with a slope of 49.5 $s^{-1}$, indicating addition of roughly 50 more action potentials per each additional second of cycle period. For all parameter sets, the driver neuron duty cycle was between 0.3057 and 0.3436 (Fig. 3C).

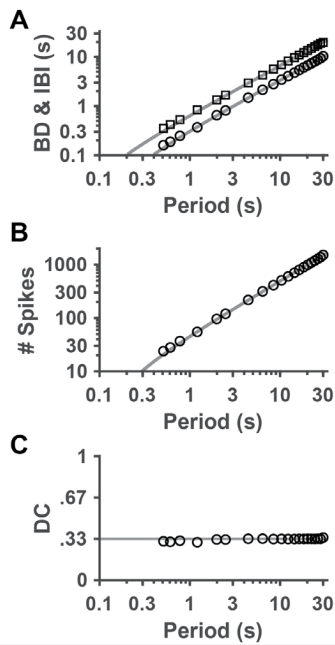

**Fig. 3 Coregulation of potassium and hyperpolarization-activated current kinetics allows maintenance of endogenously bursting driver neuron duty cycle in a range of periods. (A)** Burst duration and interburst interval of the endogenously bursting driver neuron AB plotted against pattern period on a log-log scale. The burst duration and interburst interval scale linearly with pattern period (slopes 0.33608 and 0.66392, respectively). **(B)** The number of spikes per burst plotted against pattern period on a log-log scale. The number of spikes per burst grows linearly with period and subsequently burst duration and interburst interval (slope 49.53). **(C)** The normalized duty cycle of the driver neuron is plotted against pattern period on a log-linear scale. The line corresponds to a one-third duty cycle, the ideal for bursting in a triphasic rhythm on a circle. The duty cycles are nearly constant at all periods and remain close to one-third.

The follower neurons with identical parameters exhibited similar behavior: burst duration, interburst interval and number of spikes per burst scaled linearly (Fig. 4 A-B). We computed linear regressions for the three behavioral quantities which had slopes of 0.3299, 0.6701, and 49.1 s$^{-1}$, respectively. Burst duration and interburst interval slopes differed by 0.0062 from the driver, and the number of spikes per burst slope differed by 0.43 s$^{-1}$ from the driver. This variation indicated that the follower neurons tended to have shorter burst duration, fewer spikes per burst, and longer interburst intervals than the driver neuron. Such differences are attributed to network properties. Because, in isolation, the duty cycle of the driver (and consequently the followers) is not precisely a third, no neuron bursts for exactly one-third of the pattern period. Due to the bursting order, when AB and LP burst for slightly more than a third of the period, the PY burst will be truncated by AB's next burst. This timing results in fewer spikes per PY burst and smaller burst duration overall. Additionally, timing variations arise from synaptic parameters such as time constant and activation steepness: the follower's response to inhibition is rapid but not instantaneous. Nonetheless, cellular properties determined by Mechanism 1 ensured that the behavior of the follower neurons qualitatively matched the behavior of the driver neuron. The duty cycle of both follower neurons was nearly a third across periods (Fig. 4C) and the phases of the follower neurons relative to the driver were consistent across periods, approximately uniformly dividing the interburst interval of the driver (Fig. 4D).

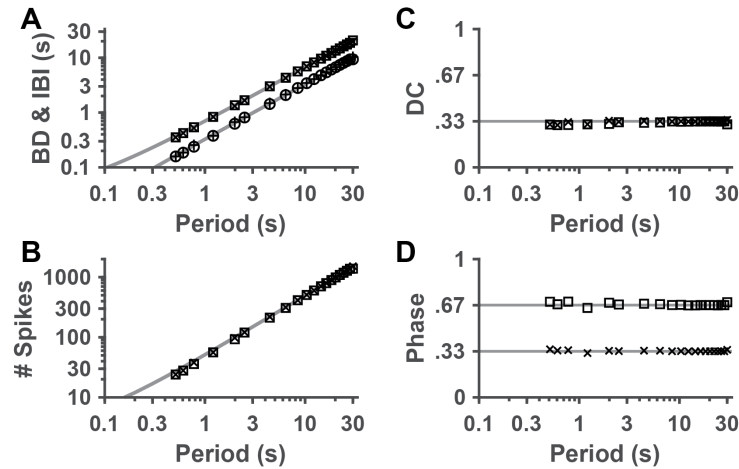

**Fig. 4 Identical coregulation of potassium and hyperpolarized-activated current kinetics in bursting driver and follower neurons supports phase maintenance in Ensemble 1. (A)** Burst duration of the follower neurons plotted against pattern period on a log-log scale where + and ○ denote LP and PY, respectively. Interburst interval of the follower neurons plotted against pattern period on a log-log scale where x and □ denote LP and PY, respectively. The burst duration and interburst interval scale linearly with pattern period (slopes 0.32987 and 0.67013, respectively). **(B)** The number of spikes per burst plotted against pattern period on a log-log scale; LP bursts and PY bursts are denoted by x and □, respectively. The number of spikes per burst grows linearly with period and subsequently burst duration and interburst interval (slope 49.095). **(C)** The normalized duty cycle of the follower neurons is plotted against pattern period on a log-linear scale where x and □ denote LP and PY, respectively. Duty cycles are nearly constant across pattern periods; the line denotes the ideal one-third duty cycle. **(D)** The normalized phase of LP and PY bursts relative to the driver AB plotted against pattern period on a log-linear scale. LP bursts and PY bursts are denoted by x and □, respectively. The lines correspond to one-third and two-thirds phase. Phase remains nearly constant for both follower neurons across periods.

In the other three ensembles, the endogenously bursting driver AB neuron used the same parameters selected in Ensemble 1. Therefore, the periods demonstrated in the other ensembles match the set collected in Ensemble 1, facilitating comparisons.

Thus, we have shown that a codimension-2 bifurcation model three-neuron pyloric motif with all bursting neurons produces a triphasic rhythm across a wide cycle period range, including the pacemaker-driven pyloric network cycle period range (~0.6 to 1.6 s; [54]) and beyond (0.5 to 30 s).

### 3.2 Phase-maintenance in Ensemble 2 where the two follower neurons are endogenously silent

In Ensemble 2, the endogenously bursting driver neuron has two endogenously silent follower neurons. Utilizing Mechanism 2, the parameters of LP and PY may be selected so that each would produce a transient burst in response to inhibitory synaptic input. When the synapses are properly weighted, the inhibition from the driver generates a triphasic bursting pattern as follows: AB bursts endogenously, inhibiting LP and PY. LP strongly inhibits PY so that LP bursts in response to the driver first, and after the conclusion of the LP burst, the PY produces a burst primarily in response to inhibition received from LP. We used the parameter sets from Mechanism 1 that give the endogenously bursting driver a duty cycle of approximately one-third over a range of periods as in Ensemble 1 (supplementary Table 1). Then, we used Mechanism 2 to coordinate the silent follower neurons such that they respond to inhibition with a single burst with one-third duty cycle at the appropriate period. Mechanism 2 controls the temporal characteristics of pulse-triggered responses in silent neurons by varying $\theta_{K2}$ towards the saddle-node bifurcation for equilibria in the regime of monostable silent regime. The transient burst is triggered by a sufficient perturbation when $\theta_h$ is near the border of the SNIC bifurcation. To place the follower neurons close to this border, we keep $\theta_h$ = -0.0415 V (Fig. 1A, blue arrow). AB was set with the parameters from Ensemble 1 that generate bursting with period 0.5 s and duty cycle approximately one-third ($\theta_h$ = 0.04123 V, $\theta_{K2}$ = -0.0041 V). The value of $\theta_{K2}$ for the follower neurons was also set to -0.0041 V. By setting identical $\theta_{K2}$ values in the driver and follower neurons, the duration of the transient bursts in the followers was set equal to the duration of the driver's endogenous bursts according to Mechanism 2. The order of follower neuron bursting was established through synaptic weighting identical to Ensemble 1 (values shown in Table 2). With the firing order established and the burst durations of each neuron all equivalent, the triphasic pattern was generated (Fig. 2 B1-B3). This procedure was repeated for all 20 cycle period parameter settings of the driver neuron (supplementary Table 1).

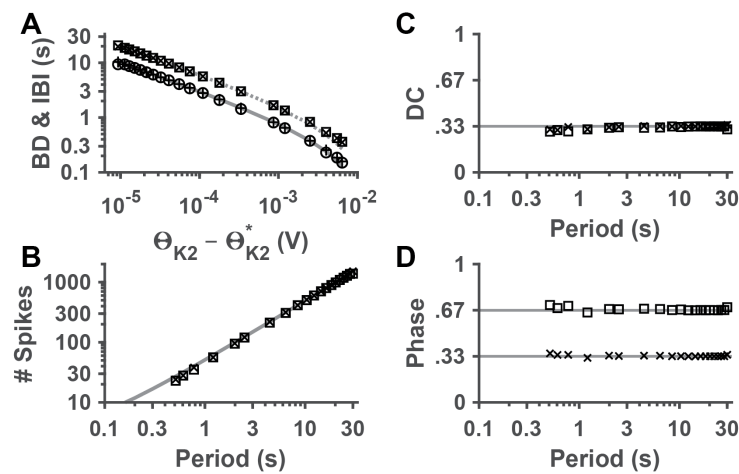

**Fig. 5 Coregulation of potassium and hyperpolarization-activated current kinetics in the driver neuron with symmetric regulation of potassium current kinetics of endogenously silent follower neurons supports phase maintenance in Ensemble 2. (A)** Burst duration and interburst interval of the endogenously silent follower neurons plotted against the distance of $\theta_{K2}$ from its critical value (-0.010505 V) on a log-log scale. Endogenously silent neurons exhibit a transient burst in response to synaptic inhibitory input from the driver neuron, which establishes a functional

periodic burst pattern. Burst duration and interburst interval of the LP neuron are denoted by + and ×, respectively. Burst duration and interburst interval of the PY neuron are denoted by ○ and □, respectively. The solid line is a curve fit for the burst duration of the LP neuron against $\theta_{K2}$ where $\theta_h$ is fixed at -0.0415 V, defined by $\frac{a}{\sqrt{|\theta_{K2}-\theta_{K2}*|}} + b$ where a = 0.03237 s, $\theta_{K2}*$ = -0.010505 V and b = 0.2705199 s. The dotted line is a curve fit for the interburst interval of the LP neuron against $\theta_{K2}$ where $\theta_h$ is fixed at -0.0415 V, defined by $\frac{a}{\sqrt{|\theta_{K2}-\theta_{K2}*|}} + b$ where a = 0.06474007 s, $\theta_{K2}*$ = -0.010505 V and b = 0.54103987 s. **(B)** The number of spikes per burst plotted against pattern period. LP bursts and PY bursts are denoted by × and □, respectively. The number of spikes per burst grows linearly with period and subsequently burst duration (slope 48.921). **(C)** The normalized duty cycle of the follower neurons is plotted against pattern period. LP and PY duty cycles are denoted by × and □, respectively. The line corresponds to a one-third duty cycle, the ideal for triphasic bursting. The duty cycles are nearly constant at all periods and remain close to one-third. **(D)** The normalized follower neuron phases relative to AB are plotted against period pattern. LP and PY phases are denoted by × and □, respectively. The lines correspond to one-third and two-thirds expected in an evenly split triphasic pattern. The phases are nearly constant at all periods and remain centered around one-third and two-thirds.

As $\theta_{K2}$ of the silent neurons approached a saddle-node bifurcation for periodic orbits, burst duration and interburst interval scaled according to the inverse-square-root law (Fig. 5A). We performed a curve fit of burst duration applied to the distance of $\theta_{K2}$ from its bifurcation value, -0.010505 V when $\theta_h$ is fixed at -0.0415 V, against burst duration at each parameter set. The curve fit for burst duration was defined by $\frac{a}{\sqrt{|\theta_{K2}-\theta_{K2}*|}} + b$ where a = 0.03237, $\theta_{K2}*$ = -0.010505 and b = 0.2705199. The curve fit for the interburst interval was defined by $\frac{a}{\sqrt{|\theta_{K2}-\theta_{K2}*|}} + b$ where a = 0.06474007, $\theta_{K2}*$ = -0.010505 and b = 0.54103987. The number of spikes per burst grew linearly with period (Fig. 5B). We computed a linear regression for the number of spikes with slope 48.9 s$^{-1}$, close to the corresponding value in Ensemble 1. This result suggests that the mechanism for increasing transient burst duration in silent neurons is also accomplished with the addition of spikes to the burst waveform. The duty cycle of the driver neuron at all parameter settings was consistent with the results of Ensemble 1 (Fig. 3C), and the duty cycles of the follower neurons at all parameter settings were also nearly constant (Fig. 5C). The duty cycles of the follower neurons ranged from 0.3059 to 0.3436 in LP and 0.2943 to 0.3326 in PY. The duty cycles of PY were lower due to the tendency of the PY burst to get truncated by the subsequent endogenous AB burst. Just as in Ensemble 1, if the duty cycle of the driver neuron were exactly a third and the follower neurons began to burst instantaneously after the driver, the duty cycle of the followers would also be a third. However, slight variance in the duty cycle of the driver combined with short gaps in firing due to synaptic properties elicit variations in the timing of bursts. Therefore, the end of the PY neuron's transient burst is sometimes suppressed by the driver, which continues its pace without any feedback from the followers. The phases of the follower neuron bursts relative to the driver neuron, measured as the time from the first spike of the driver's burst to the first spike of the follower's burst divided by pattern period, were nearly constant at all periods (Fig. 5D). The phases of the LP follower start ranged from 0.3220 to 0.3542 and the phases of the PY follower ranged from 0.6502 to 0.7040. The phases are roughly one-third and two-thirds, consistent with an approximately even split of the driver neuron's interburst interval between the followers.

Thus, we have shown that our model three-neuron pyloric motif with silent follower neurons produces a triphasic rhythm across a wide cycle period range.

### 3.3 Phase-maintenance in Ensemble 3 where the two follower neurons are endogenously spiking

In Ensemble 3, an endogenously bursting driver neuron paces two endogenously spiking follower neurons. The targeted triphasic pattern is elicited as follows: AB fires an endogenous burst which inhibits LP and PY into hyperpolarized quiescence. LP resumes spiking immediately after, and PY begins to fire after a set time equal to one preset LP burst duration. When PY resumes spiking, it inhibits LP into hyperpolarized silence. The pattern repeats when AB bursts again. If the duty cycle of the driver and the time until PY resumes spiking are properly coordinated, three equal-duration bursts partition the pattern.

The duty cycle of the driver neuron is maintained at all periods by Mechanism 1. Here, mechanism 3 is applied to control pulse-triggered responses of the endogenously spiking follower neuron PY (similarly it could be applied to LP). The mechanism is triggered by inhibition from AB's burst, which induces a pause in firing in the suppressed spiking neuron as the phase point passes through the ghost of a saddle-node bifurcation near the Cornerstone bifurcation. By selecting $\theta_h$ in the vicinity of the ghost and asymptotically approaching the border of the saddle-node bifurcation for equilibria, the latency to firing may take on an arbitrarily large value. In this ensemble, follower neurons must possess different intrinsic cellular properties to enforce the proper burst order. LP must have no delay from inhibition while PY's latency to firing must be set to equal AB's burst duration (matching the LP burst duration). To accomplish this, we first used Mechanism 1 to give AB approximately one-third duty cycle with period 0.5 s, applying the same cycle period coordinates from Ensemble 1. Using Mechanism 3, the endogenously spiking LP follower neuron was fixed with parameters sufficiently far from the Cornerstone bifurcation that there was no observable delay after inhibition ($\theta_h = 0.040$ V, $\theta_{K2} = -0.0107$ V). To situate PY near the ghost of the saddle-node bifurcation, we fixed $\theta_{K2} = -0.0106999$ V and approached the saddle-node bifurcation for equilibria with $\theta_h$ (Fig. 1A, pink arrow). We then selected $\theta_h$ for PY such that the delay was ~ 0.167 seconds, corresponding to AB's burst duration. This choice yielded the desired pattern (Fig. 2 C1-C3). We repeated the procedure for each of the AB parameter sets determined in Ensemble 1 to generate 20 patterns in a range of periods (Fig. 6; supplementary Table 1).

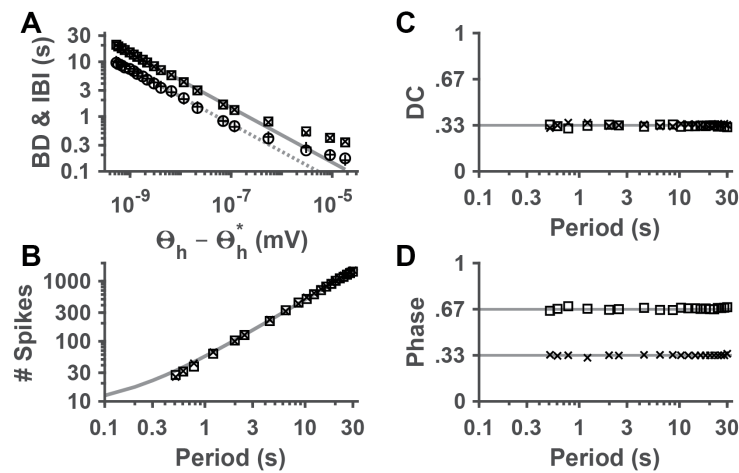

**Fig. 6 Coregulation of potassium and hyperpolarization-activated current kinetics in the driver neuron with symmetric regulation of hyperpolarization-activated current kinetics of endogenously spiking follower neurons supports phase maintenance in Ensemble 3. (A)** Burst duration and interburst interval of the endogenously spiking follower LP and PY neurons plotted against the distance of $\theta_h$ from its critical value (-0.041358046586 V) on a log-log scale. Endogenously spiking neurons exhibit a transient hyperpolarized latency to firing in response to synaptic inhibitory input from the driver neuron, which establishes a functional periodic burst pattern. Burst duration and interburst interval of the LP neuron are denoted by + and ×, respectively. Burst duration and interburst interval of the PY neuron are denoted

by ○ and □, respectively. The solid line is a curve fit for the interburst interval against $\theta_h$ of PY, where $\theta_{K2}$ is fixed at -0.0106999 V. The curve is defined by $\frac{a}{\sqrt{|\theta_h - \theta_{h*}|}} + b$ where $a = 4.608614 \times 10^{-4}$ s, $\theta_h^* = -0.041358046586$ V, and b = 0 s. The dotted line is a curve fit for the burst duration against $\theta_h$ of PY, where $\theta_{K2}$ is fixed at -0.0106999 V. The curve is defined by $\frac{a}{\sqrt{|\theta_h - \theta_{h*}|}} + b$ where $a = 2.304307 \times 10^{-4}$ s, $\theta_h^* = -0.041358046586$ V, and b = 0 s. **(B)** The number of spikes per burst plotted against pattern period. LP bursts and PY bursts are denoted by × and □, respectively. The number of spikes per burst grows linearly with period and subsequently burst duration (slope 48.699). **(C)** The normalized duty cycle of the follower neurons is plotted against pattern period. LP and PY duty cycles are denoted by × and □, respectively. The line corresponds to a one-third duty cycle, the ideal for triphasic bursting. The duty cycles are nearly constant at all periods and remain close to one-third. **(D)** The normalized follower neuron phases relative to AB are plotted against period pattern. LP and PY phases are denoted by × and □, respectively. The lines correspond to one-third and two-thirds phases expected in an evenly split triphasic pattern. The phases are nearly constant at all periods and remain centered around one-third and two-thirds.

In this ensemble, we found that the synaptic parameters of Ensembles 1 and 2 prevented the proper function of Mechanism 3 such that delay would not grow according to the inverse-square-root law and was not observed for more than a few seconds. We restored the mechanisms' functionality by adjusting the synaptic reversal potential closer to the stable hyperpolarized equilibria (-0.041358046586 V). A second problem arose as soon as we simulated the full circuit: the LP follower's inhibition onto PY blocked the mechanism. Strength of LP to PY synaptic inhibition had to be zero or very small, otherwise PY would stay silent for the predetermined PY window interval. All presented data were obtained with this synaptic strength set to zero, but the appropriate pattern would be generated with a synaptic strength up to 2 x $10^{-8}$ nS at all cycle periods. It is of note that in the living pyloric network, the LP to PY synapse is demonstrated to have no significant effect on the temporal characteristics of PY, and its removal does not hinder PY's ability to maintain phase with the driver [55]. We strengthened the synapse from PY to LP to ensure that when PY returned to spiking, LP was suppressed into a hyperpolarized state. This mechanism was not required when the duration of LP's bursting activity was fixed (Ensembles 1 and 2).

Unlike Ensembles 1 and 2, the firing waveform of the follower neurons was not identical to that of the endogenously bursting driver. A follower neuron burst did not develop the increasingly depolarized spikes near termination that are characteristic of bursts established with Mechanisms 1 and 2. Instead, the waveform of every spike beyond the first was nearly identical. This similarity occurred because when the follower neurons are active, they exhibit oscillations near the spiking limit cycle, and what we refer to as a "burst" is a fixed interval of spiking terminated with a synaptic input which determines the duration of the following pause (i.e., interburst interval).

As $\theta_h$ of the endogenously spiking follower PY approached the saddle-node bifurcation for equilibria, interburst interval and burst duration scaled according to the inverse-square-root law (Fig. 6A). We performed a curve fit applied to the distance of $\theta_h$ from its bifurcation value, -0.041358046586 V when $\theta_h$ is fixed at -0.0106999 V, against interburst interval at each parameter set. The curve fit was defined by $\frac{a}{\sqrt{|\theta_h - \theta_{h*}|}} + b$ where $a = 4.608614 \times 10^{-4}$, $\theta_h^* = -0.041358046586$, and b = 0. The curve fit for the burst duration was defined by $\frac{a}{\sqrt{|\theta_h - \theta_{h*}|}} + b$ where $a = 2.304307 \times 10^{-4}$, $\theta_h^*$ = -0.041358046586, and b = 0. The number of spikes per burst grew linearly with period (Fig. 6B). We computed a linear regression for the number of spikes with slope 48.7 $s^{-1}$, close to the corresponding values in Ensembles 1 and 2. This result suggests that the addition of ~ 50 spikes per additional second of cycle period corresponds to the spiking periodic orbit in the region of monostable spiking, similar to Ensembles 1 and 2.

The duty cycle of the driver neuron at all parameter settings was consistent with the results of Ensemble 1 (Fig. 4C), and the duty cycles of the follower neurons at all parameter settings were also nearly constant (Fig. 4C). The duty cycles of the follower neurons ranged from 0.3129 to 0.3526 in LP and 0.3105 to 0.3389 in PY. The phases of the follower neuron bursts to the driver neuron, measured as the time from the first spike of the driver's burst to the first spike of the follower's burst divided by pattern period, were nearly constant at all periods (Fig. 4D). The phases of the LP follower ranged from 0.3154 to 0.3445 and the phases of the PY follower ranged from 0.6554 to 0.6881. The phases are roughly one-third and two-thirds, consistent with an approximately even split of the driver neuron's interburst interval between the followers.

Thus, we have shown that our model three-neuron pyloric motif with spiking follower neurons produces a triphasic rhythm across a wide cycle period range.

### 3.4   Phase-maintenance in Ensemble 4 which includes one silent follower neuron and one spiking follower neuron

In Ensemble 4, an endogenously bursting driver has an endogenously silent follower and an endogenously spiking follower. With these settings, the desired pattern is generated as follows: AB bursts endogenously, immediately after which the LP neuron fires a burst of equal duration in response to inhibition. Simultaneously, the PY neuron is suppressed into hyperpolarization by the inhibition from the driver. The PY neuron's latency to firing is equal to the sum of the burst durations of the AB driver and the LP follower; therefore, it resumes spiking at the same time as the LP burst terminates. Thus, the inhibition-triggered responses and their controlling mechanisms from the two previous ensembles work in conjunction to generate a qualitatively identical pattern. We achieved the pattern by coordinating with all three mechanisms. The driver was controlled through Mechanism 1, the endogenously silent follower neuron was controlled by Mechanism 2, and the endogenously spiking follower neuron was controlled by Mechanism 3. For all settings of the driver neuron, the LP follower neuron was selected symmetrically across the border of the SNIC bifurcation to have identical burst duration in response to inhibitory synaptic input. The PY follower neuron was selected in the vicinity of the ghost of the saddle-node bifurcation such that it exhibited latency to firing in response to synaptic input with duration equal to that of the driver's burst. The synaptic parameters from Ensemble 3 were continued for the AB to PY synapse, while the Ensemble 1 and 2 synaptic parameters were continued for the AB to LP synapse.

We first considered the AB parameters that yield bursting with one-third duty cycle with period 0.5 s. These coordinates were taken from Ensemble 1. We selected the corresponding LP parameters from Ensemble 2, which ensured that LP fired in response to AB with a burst approximately 167 ms long. Finally, we selected the corresponding PY parameters from Ensemble 3, which fixed the delay to firing in response to AB to approximately 167 ms. This series of selection generated the desired pattern (Fig. 2 D1-D3). We repeated the procedure for all 20 sets of AB cycle period parameters (Fig. 7; supplementary Table 1). The burst duration and interburst interval of the silent LP neuron grew identically to that of Ensemble 2, and the burst duration and interburst interval of the endogenously spiking PY neuron grew identically to that of Ensemble 3 (Fig. 7A and 7B, respectively). The same curve fits were applied. The number of spikes per burst grew linearly with period (Fig. 7C). We computed a linear regression for the number of spikes with slope 48.699, which is equal to the slope computed for Ensemble 3. The duty cycle of the driver neuron at all parameter settings was consistent with the results of Ensemble 1 (Fig. 3C), and the duty cycles of the follower neurons at all parameter settings were also nearly constant (Fig. 7D). The duty cycles of the follower neurons ranged from 0.2943 to 0.3523 in LP and 0.3105 to 0.3389 in PY. The phases of the follower neuron bursts to the driver neuron, measured as the time from the first spike of the driver's burst to the first spike of the follower's burst divided by pattern period, were nearly constant at all periods (Fig. 7E). The phases of LP ranged from 0.3162 to 0.3541 and the phases of PY ranged from 0.6554 to 0.6881. The phases were roughly one-third and two-thirds, consistent with an approximately even split of the driver neuron's interburst interval between the followers. Note that the phases and duty cycles of the PY neuron at all periods exactly matched the values in Ensemble 3; this result was expected according to the function of Mechanism 3.

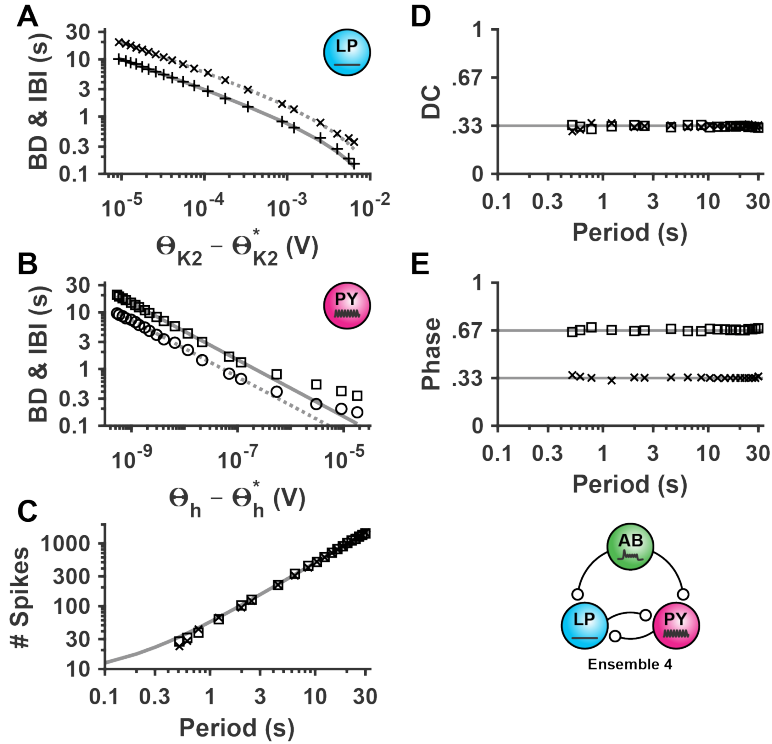

**Fig. 7 Coordinated control of potassium current kinetics and hyperpolarization-activated current kinetics in endogenously silent and spiking follower neurons supports phase maintenance in Ensemble 4. (A)** Burst duration and interburst interval of the endogenously silent follower neuron LP plotted against the distance of $\theta_{K2}$ from its critical value (-0.010505 V) on a log-log scale. Endogenously silent neurons exhibit a transient burst in response to synaptic inhibitory input from the driver neuron, which establishes a functional periodic burst pattern. Burst duration and interburst interval of the LP neuron are denoted by + and ×, respectively. The solid line is a curve fit for the burst duration of the LP neuron against $\theta_{K2}$ where $\theta_h$ is fixed at -0.0415 V, defined by $\frac{a}{\sqrt{|\theta_{K2}-\theta_{K2}*|}} + b$ where a = 0.03237 s, $\theta_{K2}*$ = -0.010505 V and b = 0.2705199 s. The dotted line is a curve fit for the interburst interval of the LP neuron against $\theta_{K2}$ where $\theta_h$ is fixed at -0.0415 V, defined by $\frac{a}{\sqrt{|\theta_{K2}-\theta_{K2}*|}} + b$ where a = 0.06474007 s, $\theta_{K2}*$ = -0.010505 V and b = 0.54103987 s. **(B)** Burst duration and interburst interval of the endogenously spiking follower PY neuron is plotted against the distance of $\theta_h$ from its critical value (-0.041358046586 V) on a log-log scale. Endogenously spiking neurons exhibit a transient hyperpolarized latency to firing in response to synaptic inhibitory input from the driver neuron, which establishes a functional periodic burst pattern. Burst duration and interburst interval of the PY neuron are denoted by ○ and □, respectively. The solid line is a curve fit for the interburst interval against $\theta_h$ of PY, where $\theta_{K2}$ is fixed at -0.0106999 V. The curve is defined by $\frac{a}{\sqrt{|\theta_h-\theta_h*|}} + b$ where a= 4.608614 x $10^{-4}$, $\theta_h*$ = -0.041358046586 V, and b = 0 s. The dotted line is a curve fit for the burst duration against $\theta_h$ of PY, where $\theta_{K2}$ is fixed at -0.0106999 V. The curve is defined by $\frac{a}{\sqrt{|\theta_h-\theta_h*|}} + b$ where a= 2.304307 x $10^{-4}$, $\theta_h*$ = -0.041358046586 V, and b = 0 s. **(C)** The number of spikes per burst plotted against pattern period. LP bursts and PY bursts are denoted by × and □, respectively. The number of spikes per burst grows linearly with period and subsequently burst duration (slope 48.699). **(D)** The normalized duty cycle of the follower neurons is plotted against pattern period. LP and PY duty cycles are denoted by × and □, respectively. The line corresponds to a one-third duty cycle, the ideal for triphasic bursting. The duty cycles are nearly

constant at all periods and remain close to one-third. **(E)** The normalized follower neuron phases relative to AB are plotted against period pattern. LP and PY phases are denoted by × and □, respectively. The lines correspond to one-third and two-thirds phases expected in an evenly split triphasic pattern. The phases are nearly constant at all periods and remain centered around one-third and two-thirds.

Thus, we have shown that our model three-neuron pyloric motif with heterogenous follower neurons produces a triphasic rhythm across a wide cycle period range.

## 3.5   Variation of three parameters is sufficient for triphasic rhythm production in most network ensembles

To summarize our results thus far, we have built a three-neuron network inspired by the pyloric CPG that is able to produce a triphasic rhythm with a variety of follower neuron type combinations (Ensembles 1-4). Tuning this network to produce functional motor patterns across a wide range of cycle periods can require simultaneous adjustments of as many as six parameters (supplementary Table 1; $\theta_h$ & $\theta_{K2}$ in each of three neurons). While it is known that neuromodulatory inputs to CPGs can regulate neuronal conductances and firing properties in common [57, 58], the feasibility of so many parameters being tuned simultaneously in a correct manner seems unlikely or would not be robust. Thus, we investigated the minimal amount of neuronal parameter variation in the network that would still produce a triphasic rhythm across cycle periods.

As we have shown before, the model pacemaker AB neuron can be made to vary its cycle period over a wide range while maintaining approximately one third duty cycle (Figs. 2-7). This behavior requires covarying two AB neuron parameters ($\theta_h$ and $\theta_{K2}$) to scale both interburst interval and burst duration. Prior work in our lab has shown how $\theta_{K2}$ variation can control burst duration and $\theta_h$ regulates delay to firing or interburst interval [48]. We were unable to produce a triphasic motor pattern in any of the ensembles across all tested cycle periods while keeping all follower neuron $\theta_h$ and $\theta_{K2}$ parameters constant (data not shown). Thus, we sought to investigate the minimum variation in follower neuron parameters that could still produce a triphasic motor pattern. To this end, we started with the working triphasic follower neuron parameters for the longest cycle period (30 s) for all four ensembles and attempted to keep as many of these follower neuron parameters constant as possible while maintaining phasing at all shorter cycle periods. The network ensembles where three parameter variations (two of which are the AB's $\theta_h$ and $\theta_{K2}$; the third is in one of the follower neurons) is sufficient for triphasic rhythm production are described below.

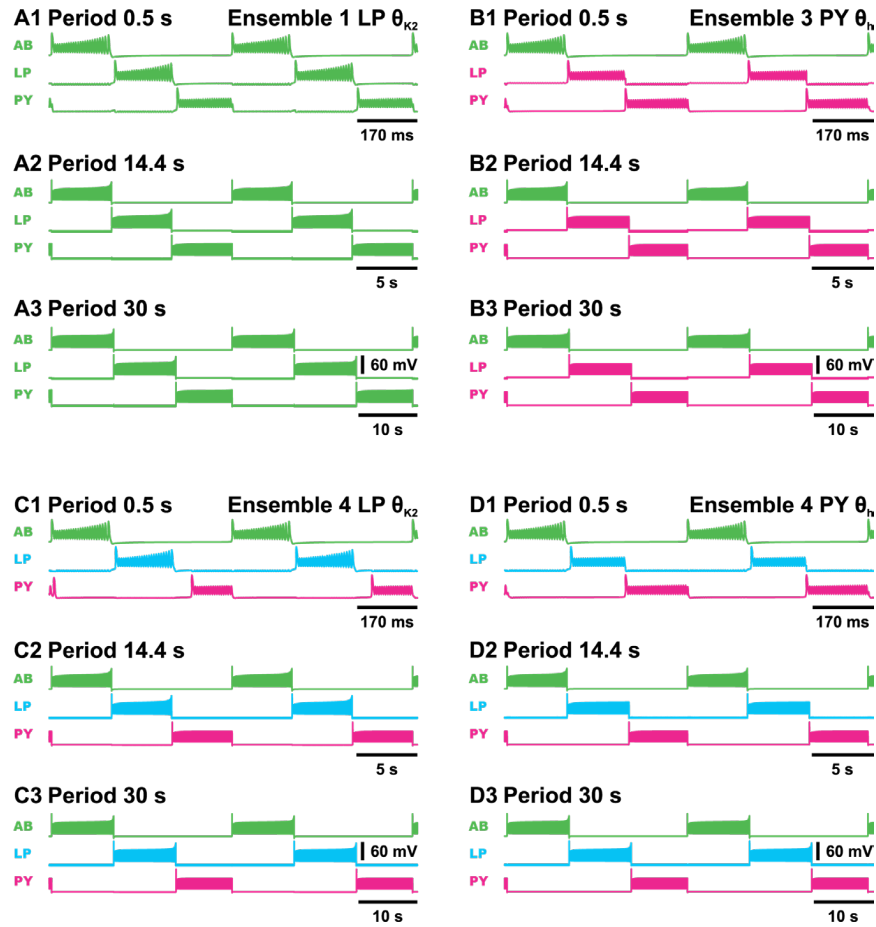

**Fig. 8 Variation of only three parameters is sufficient for triphasic rhythms in ensembles with endogenously bursting follower neurons (Ensemble 1), endogenously spiking follower neurons (Ensemble 3), and with one endogenously silent follower and one endogenously spiking follower neuron (Ensemble 4).** The patterns are shown at periods 0.5 s, 14.4 s, and 30 s. Panels **A-D** include a shared variation of two of the parameters ($\theta_h$ and $\theta_{K2}$) in the AB neuron to set the network period. The third parameter varied depends on which Ensemble is configured and which neuron is regulated (LP: $\theta_{K2}$; PY: $\theta_h$). Panel **A** comprises an Ensemble 1 network with LP neuron $\theta_{K2}$ variation. Coordinates for all neurons in **(A1)** are (AB: $\theta_h = 0.04123$ V, $\theta_{K2} = -0.0041$ V; LP: $\theta_h = 0.04135646145$ V, $\theta_{K2} = -0.0041$ V; PY: $\theta_h = 0.04135646145$ V, $\theta_{K2} = -0.010496$ V), coordinates for all neurons in **(A2)** are (AB: $\theta_h = 0.04135621058$ V, $\theta_{K2} = -0.0104644$ V; LP: $\theta_h = 0.04135646145$ V, $\theta_{K2} = -0.0104644$ V; PY: $\theta_h = 0.04135646145$ V, $\theta_{K2} = -0.0104644$ V), and coordinates for all neurons in **(A3)** are (AB: $\theta_h = 0.04135646145$ V; PY: $\theta_h = -0.010496$ V; LP: $\theta_h = 0.04135646145$ V, $\theta_{K2} = -0.010496$ V; PY: $\theta_h = 0.04135646145$ V, $\theta_{K2} = -0.010496$ V). Panel **B** comprises an Ensemble 3 network with PY neuron $\theta_h$ variation. Coordinates for **(B1)** are (AB: $\theta_h = 0.04123$ V, $\theta_{K2} = -0.0041$ V; LP: $\theta_h = 0.04$ V, $\theta_{K2} = -0.0107$ V; PY: $\theta_h = 0.04134$ V, $\theta_{K2} = -0.0106999$ V). Coordinates for **(B2)** are (AB: $\theta_h = 0.04135621058$ V, $\theta_{K2} = -0.0104644$ V; LP: $\theta_h = 0.04$ V, $\theta_{K2} = -0.0107$ V; PY: $\theta_h = 0.0413580444$ V, $\theta_{K2} = -0.0106999$ V). Coordinates for **(B3)** are (AB: $\theta_h = 0.04135646145$ V, $\theta_{K2} = -0.010496$ V; LP: $\theta_h = 0.04$ V, $\theta_{K2} = -0.0107$ V; PY: $\theta_h = 0.04135804605$ V, $\theta_{K2} = -0.0106999$ V). Panel **C** comprises an Ensemble 4 network with LP neuron $\theta_{K2}$ variation. Coordinates for **(C1)** are (AB: $\theta_h = 0.04123$ V, $\theta_{K2} = -0.0041$ V; LP: $\theta_h = 0.0415$ V, $\theta_{K2} = -0.0041$ V; PY: $\theta_h = 0.04135804605$ V, $\theta_{K2} = -0.0106999$ V). Coordinates for **(C2)** are (AB: $\theta_h = 0.04135621058$ V, $\theta_{K2} = -0.0104644$ V; LP: $\theta_h = 0.0415$ V, $\theta_{K2} = -0.0104644$ V; PY: $\theta_h = 0.04135804605$ V, $\theta_{K2} = -0.010699$ V). Coordinates for **(C3)** are (AB: $\theta_h = 0.04135646145$ V, $\theta_{K2} = -0.010496$ V; LP: $\theta_h = 0.0415$ V, $\theta_{K2} = -0.010496$ V; PY: $\theta_h = 0.04135804605$ V, $\theta_{K2} = -0.0106999$ V). Panel **D** comprises an Ensemble 4 network with PY neuron $\theta_h$ variation. Coordinates for **(D1)** are (AB: $\theta_h = 0.04123$ V, $\theta_{K2} = -0.0041$ V; LP: $\theta_h = 0.0415$ V, $\theta_{K2} = -0.010496$ V; PY: $\theta_h = 0.04134$ V, $\theta_{K2} = -0.0106999$ V). Coordinates for **(D2)** are (AB: $\theta_h = 0.04135621058$ V, $\theta_{K2} = -0.0104644$ V; LP: $\theta_h = 0.0415$ V, $\theta_{K2} = -0.010496$ V; PY: $\theta_h = 0.041358044$ V, $\theta_{K2} = -0.0106999$ V). Coordinates for **(D3)** are (AB: $\theta_h = 0.04135646145$ V, $\theta_{K2} = -0.010496$ V; LP: $\theta_h = 0.0415$ V, $\theta_{K2} = -0.010496$ V; PY: $\theta_h = 0.04135804605$ V, $\theta_{K2} = -0.0106999$ V).

In Ensemble 1, all three neurons are endogenous bursters. By varying LP burst duration via $\theta_{K2}$, we were able to keep LP duty cycle at approximately one third across cycle periods and allow the PY neuron to also maintain one third duty cycle before being shut off by the next cycle of AB neuron activity (Fig. 8 A1-A3). In Ensemble 2, both follower neurons are endogenously silent. We were unable to produce a triphasic rhythm at all cycle periods with a single follower neuron parameter variation (data not shown). In Ensemble 3, both follower neurons are endogenously spiking. By varying PY delay to firing via $\theta_h$, we were able to allow for a PY delay to firing of one third duty cycle. When the PY neuron began to burst, it would shut off the LP neuron leading to one third LP duty cycle. The PY neuron would then be shut off at one third duty cycle by the next cycle of AB neuron activity (Fig. 8 B1-B3). In Ensemble 4, an endogenously bursting AB neuron has an endogenously silent follower LP neuron and an endogenously spiking follower PY neuron. Both described strategies above (LP $\theta_{K2}$ burst duration and PY $\theta_h$ delay to firing) were separately applied to Ensemble 4. Regulation of LP neuron burst duration produced triphasic rhythms across cycle periods (Fig. 8 C1-C3), although the PY neuron was delayed at the shortest cycle period leading to a reduced duty cycle. Regulation of PY neuron delay to firing produced triphasic rhythms across cycle periods (Fig. 8 D1-D3). These functional neuron parameters are shown in supplementary Table 2.

Thus, we have shown that Ensembles 1, 3, and 4 of our model three-neuron pyloric motif requires only two pacemaker neuron parameters (i.e., tuning period) and one follower neuron parameter to be tuned across cycle periods to produce a functional triphasic rhythm. It is possible that Ensemble 2 could similarly be tuned, but we were unable to find those parameter sets. All ensembles could be tuned to remain triphasic with four neuron parameters (two AB, one LP, one PY; data not shown).

## Coregulation of $I_{K2}$ and $I_h$

Since coregulation of ion currents is hypothesized to be an important contributor to phase maintenance [27, 28, 35, 38, 59], we described the phase maintenance mechanisms in terms of coordinated changes of voltages of half-activation of the coregulated currents, the potassium K2 and hyperpolarization activated currents. To compare how these changes of the activation kinetics affected currents, we represented this coregulation in a manner similar to the experimental coregulation results demonstrated by Khorkova and Golowasch in the pyloric system [59]. The steady state values of $I_{K2,\infty}(V_{IBI}, \theta_{K2})$ and $I_{h,\infty}(V_{BD}, \theta_h)$ were computed as functions of the 20 table values of $\theta_{K2}$ and $\theta_h$, and two characteristic values of membrane potential, representing burst, $V_B$, and interburst interval, $V_{IB}$, respectively, according to the expressions ($I_{K2,\infty}(V_{IBI}, \theta_{K2}) = \bar{g}_{K2}m_{K2,\infty}^2(V_{IBI}, \theta_{K2})[V_{IBI} - E_K]$ and $I_{h,\infty}(V_{BD}, \theta_h) = \bar{g}_h m_{h,\infty}^2(V_{BD}, \theta_h)[V_{BD} - E_h]$). The two average membrane potentials <V> were calculated for the pattern with the longest period using the first five steady state periods within the trajectory. The activity was considered at its steady state when the successive cycle period durations differed by less than .1 seconds. The value $V_B$ was calculated as the average V over the middle 20 % of the burst interval which starts at the first spike and ends on the last spike. The value $V_{IB}$ interburst interval segment's endpoints were isolated by finding the 75 % mark of the interburst interval to reach a stable point in $I_h$ and identifying points that were 10 % of the interburst duration from that mark. These 20 % (65 to 85 % of IBI) were selected because they allow us to remove any interference caused by the transition between the interburst and burst intervals, and between the two modes of activity caused by different synaptic current inputs within the interburst interval for the LP and PY neurons (Fig. 9).

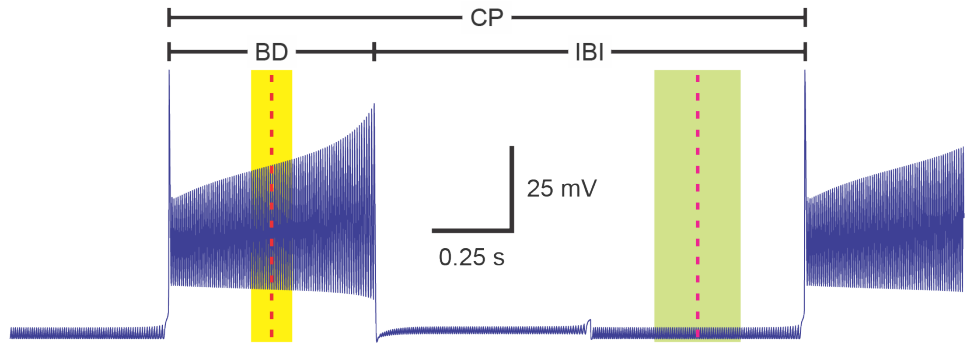

**Fig. 9 Measured regions of a single isolated period of the PY neuron.** The voltage recording example shown is from Ensemble 2 with a 1.97 s CP. Start and end of the cycle period are signified by the CP brackets. The ranges of the burst and interburst interval segments are signified by BD and IBI brackets, respectively. The selected portions of the BD and IBI for steady state current calculations are shown as yellow and green regions. In the burst duration is a red dashed line indicating the burst's 50 % mark and a yellow zone that corresponds with 40 to 60 % of BD. Inside the interburst interval is a purple dashed line indicating the IBI's 75 % mark and a green zone that corresponds with 65 to 85 % of IBI.

To limit the impact of inter-period voltage variations on calculated average steady state currents (especially at shorter CP's where one action potential more or less can cause shifts in average voltage), an average voltage was measured for the BD and IBI of the longest cycle period trajectory (30 s) for five consecutive stable cycles. These voltages were then averaged and used for calculating ion currents at all cycle periods, effectively removing noise from the voltages used in all current calculations across cycle periods.

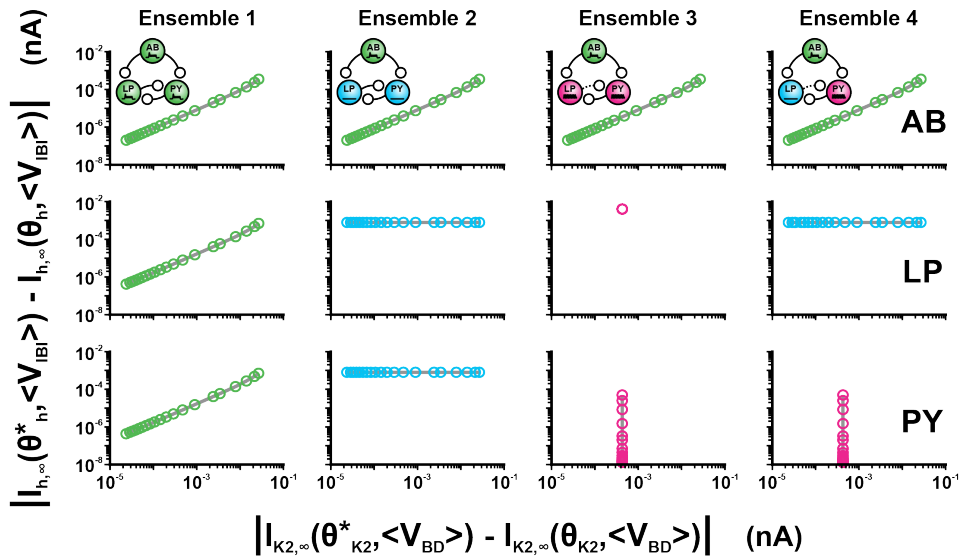

**Fig. 10 Coregulation of steady state $I_h$ and $I_{K2}$ currents across cycle periods.** The relative steady state $I_h$ and $I_{K2}$ currents were calculated for all 20 cycle periods. The $I_{K2}$ steady state current measured during the burst duration varies along the x-axis, while the $I_h$ current during the interburst interval changes along the y-axis. For all graphs, the scale is logarithmic, and the points plotted are the distance of the current value from the current of the longest cycle period (30 s; top-left corner points). All points are shifted relative to the currents from the longest cycle period: $10^{-7}$ for the potassium current and $10^{-9}$ for the h current.

In Figure 10, the coregulation of the potassium and hyperpolarization activated steady state currents in the burst duration and interburst intervals respectively can be observed. In this case, the $\theta_{K2}$ and $\theta_h$ values (through their impact on ion conductances) are the only influences on current, due to the voltage remaining constant because a single voltage from the longest cycle period was used for all the current calculations. As a result, the calculated currents change proportionally to the change in those two variables.

In Ensemble 1, all neurons have identical dynamics. Both the average $I_{K2}$ and $I_h$ decrease with the increase in period. This behavior corresponds to that of three identically bursting neurons in which both $\theta_{K2}$ and $\theta_h$ approach the bifurcation value. In Ensemble 2, the LP and PY neurons are silent, and their $\theta_h$ values remain constant. As a result, $I_h$ remains constant. The only change is that at longer cycle periods there is a decrease in $I_{K2}$ for the LP and PY neurons. In Ensemble 3, the tonic spiking LP neuron's bursting activity is fully sculpted by the AB and PY neuron inhibition because its $\theta_{K2}$ and $\theta_h$ variables do not change, thus resulting in unchanging $I_{K2}$ and $I_h$. Only $\theta_h$ changes for the PY neuron, regulating the interburst interval resulting in the decreasing $I_h$ current seen on the graph seen at longer periods. Finally, in Ensemble 4, the AB neuron is bursting, the LP neuron is silent, and the PY neuron is tonically spiking. This corresponds to $\theta_{K2}$ and $\theta_h$ being altered in the AB neuron, only $\theta_{K2}$ changing for the LP neuron, and only $\theta_h$ changing for the PY neuron. These alterations in the $\theta_{K2}$ and $\theta_h$ variables reflect on both $I_{K2}$ and $I_h$. In the AB neuron, both currents decrease with longer periods. In the LP neuron, only $I_{K2}$ decreases. Finally, in the PY neuron, only $I_h$ decreases.

From Figure 10, the function of the three mechanisms for precise temporal control in our model neurons becomes clear. In Ensemble 1, mechanism 1 is in control of the follower neurons. As $\theta_{K2}$ and $\theta_h$ approach their bifurcation value, the follower neurons $I_{K2}$ and $I_h$ currents become smaller, resulting in a longer burst durations and interburst intervals. In Ensemble 2, the increase in induced LP neuron burst duration can be seen together with a decrease in $I_{K2}$, thus demonstrating the function of mechanism 2. In Ensemble 3, mechanism 3 is seen in action in the PY neuron, where $\theta_h$ decreases together with the $I_h$ current's magnitude, while the interburst interval's duration is increased. Finally, in Ensemble 4, a coexistence of all three mechanisms is demonstrated. The AB neuron is regulated by mechanism one, while the LP neuron is regulated by mechanism 2, and the PY neuron is regulated by mechanism 3. In all three cases, the expected decrease of current size with the increase of period duration occurs as is dictated by the three mechanisms.

## 4    Discussion

For animals to survive the challenges of their environment, they must be able to adjust their behaviors, for example a lamprey that can only swim at one speed is easy prey. One way for an animal's motor behaviors to meet this demand is by maintaining the phases of motor pattern activity across cycle periods. Phase maintenance is a notable quality of many central pattern generators [60]. CPGs are tuned to support functional output through neuromodulation that orchestrates changes in membrane and synaptic properties. Many CPGs include some neurons that play a special role as a pacemaker, setting the pace of a rhythmic motor pattern. There is a multitude of experimental and computational evidence that cellular mechanisms contribute significantly to the temporal characteristics of neurons that enable their participation in carefully timed patterns. We have demonstrated a family of candidate mechanisms at work in a generic CPG model. Through the coordinated variation of two biophysical parameters, the voltage of half-activation of the non-inactivating $K^+$ current ($I_{K2}$) and hyperpolarization-activated current ($I_h$) channels, a triphasic pattern is achieved yielding phase maintenance with sixty-fold variation in period. These results are achieved in ensembles where the follower neurons exhibit three distinct endogenous activities, as well as an ensemble where the follower neurons are behaviorally heterogeneous.

## 4.1 Comparison to other mechanisms

A host of cellular and synaptic mechanisms for phase maintenance in the pyloric network have been described in both experimental and computational modeling studies.

First, Reyes et al. discovered that the relationship of oscillation period and phase lag is a dynamical invariant through dynamic clamp synaptic modification between the pacemaker PD (electrically coupled to AB) and LP, the singular feedback synapse to the pacemaker group [61]. This work does not identify a mechanism through which the synapse would directly affect phase, although it suggests that rebound dynamics are affected by the amount of synaptic inhibition. A key observation of the study is that across preparations, burst period, and phase varied significantly while the relationship of the two was tightly conserved within each preparation. The mechanisms we describe would support this diversity: we demonstrated phase maintenance in vastly different ensembles of neurons, and though we selected the conservation of approximately one-third and two-third follower neuron phases, the mechanisms should support any other division of burst activity.

## 4.2 Comparison of mechanisms to experimental results.

Our mechanism for increasing burst duration is consistent with recordings of phase maintaining driver and follower pyloric neurons, in which the number of spikes per burst and burst duration both scale with cycle period [45]. The mechanism presented here for increased burst duration functions through the addition of stereotyped spikes to the bursting waveform. The added spikes have characteristic shape, amplitude, spike width, and interspike interval. In all ensembles, we observed that the number of spikes per burst scaled linearly with cycle period at a near constant rate. There may be small-scale differences in interspike interval in the parameter space as burst duration grows but the predominant mechanism for increasing burst duration is the addition of discrete, stereotyped spikes.

To identify a neuron that may be utilizing these mechanisms, experimentalists might look for bursts that increase duration through the addition of spikes. This criterion is still relevant if the waveform of the follower neuron's activity resembles tonic spiking more than a burst, which may be characterized by a plateau potential or high amplitude initial spike. They might also look for latency to firing in a spiking neuron mediated by a slowly conducting current. The follower neurons may be conditional bursters: they participate in a firing pattern in the connected circuit but cannot maintain the activity (i.e., cycle period and duty cycle) in isolation. Experimentally, these neurons would not be endogenously bursting, but their biophysical parameters situate them near enough to the bursting regime. Our family of mechanisms explains all possible orchestrations of the motif when the Cornerstone bifurcation determines neuronal activity. The family of mechanisms around the Cornerstone bifurcation corresponds to situations where follower neurons in isolation are bursting, silent, or spiking.

While the mechanisms support phase maintenance at arbitrarily large periods, the observed period of central pattern generators is constrained to a range, as homeostasis would demand. Here, we apply the mechanisms in periods ranging from 0.5 to 30 s as a proof of principle. Boundaries on the period of a central pattern generator may be imposed not directly by the cells but by the network configuration: follower neurons can constrain the range of periods in which the pyloric network generates a functional pattern with chemical and electrical synapses to the pacemaker group [54]. Therefore, despite the capabilities of each cell individually to exhibit bursts and periods of quiescence with near limitless durations, it is likely that network properties combined with neuromodulation prevent such time scales from being realized.

The present report demonstrates how the modulation of two biophysical parameters, $\theta_h$ and $\theta_{K2}$, may support phase maintenance. Varying these parameters represents modulation of the potassium and hyperpolarization-activated currents. Hyperpolarization-activated and potassium currents are common targets for neuromodulators [29], such as dopamine [62-64], serotonin [65], and myomodulin [66, 67]. Maximum conductances and voltage of half-activation of ionic currents are biophysical parameters that are affected by neuromodulation, in particular.

Secondly, Mouser et al. demonstrated that a form of short-term synaptic plasticity called synaptic depression supports phase maintenance in an inhibitory pyloric motif identical to the present study [24]. In a model network consisting of simplified, non-spiking neurons, synaptic depression exerted increasing influence on the phase of the follower neurons at higher periods of the driver. Cellular properties predominate at shorter periods and the transient potassium current is suggested to work synergistically with synaptic depression throughout. This particular synaptic mechanism extends the range of phase maintenance through its effects on the reciprocal follower neuron inhibitory synapses, which provide a target for experimentation. It is likely that mechanisms for phase maintenance in the living system possess a combination of both cellular and synaptic mechanisms.

Thirdly, Hooper et al. determined that model pyloric neurons more closely recreate experimental phase maintenance observations when they contain a variable slowly conducting intrinsic current [27]. As in our model, their modulated parameter is the voltage of half-activation. Hooper et al. restrict modulation to the potassium current, while noting that the mechanism is non-species specific. In this cellular mechanism, the dynamics of a slowly conducting current alter a postsynaptic neuron's post-inhibition latency to firing as the current's voltage of half-activation shifts. When the sign of the potassium current's voltage dependency was switched in the model, the follower neuron PY behaved in an anti-phase maintaining manner, a potential target for dynamic clamp experiments to confirm the mechanism in the living system.

Finally, the theme of coregulated intrinsic currents presented here has been examined with statistical [28] and experimental [29, 59] methods. In the single neuron model of Soofi et al., the conductances of eight currents were varied in a pairwise fashion and correlated with the degree of spike phase maintenance in a single bursting neuron. Two pairs were positively correlated: sodium and calcium, and slow calcium and calcium-dependent potassium. This result does not match the experimental findings of MacLean, Zhang [29] nor Khorkova and Golowasch [59]. Their studies suggest that the hyperpolarization-activated and potassium (A-type) currents are coregulated to maintain spikes-per-burst and burst period in homeostasis or the maintain functional motor patterns, respectively. The Soofi et al. study deals with a qualitatively different model: a single cell, possibly acting outside the range that would activate the hyperpolarization-activated current. The coregulated pairs suggested could be examined through concurrent targeted neuromodulation in the living pyloric network.

It should be noted that the two currents in our model, $I_{K2}$ and $I_h$, are similar to those investigated in these experimental studies [29, 59]. All three studies argue for a central role for the hyperpolarization-activated current in control of neuronal activity. Instead of the relatively slow non-inactivating ($I_{K2}$) potassium current in our model that builds during a burst, both experimental studies argued for the coregulation of the fast, transient potassium (A-type) current with the hyperpolarization-activated current. Future studies will need to explore the response of our model with A-type currents to see if it can provide similar outward current effects, but it is unlikely to be as robust in its phase maintaining properties due to its speed and inactivation.

## 4.3    Choice of bifurcation parameter

Our choice of bifurcation parameters is motivated by experimental results that the expression of the hyperpolarization-activated current and potassium current channels orchestrate changes to maintain electrical identity in pyloric neurons [29, 30, 68]. A common parameter choice for coregulation analysis is the maximum conductance of ionic currents [28, 68]. The voltage of half-activation is qualitatively different as it changes ion channel kinetics and membrane dynamics. For studies of phase maintenance on the bursting activity scale, slow conductance dynamics present an intuitive and easily modulated biophysical parameter. The voltage of half-activation is not a commonly used bifurcation parameter in studies of coregulated currents. Voltage of half-activation was considered by Hooper, Buchman [27] in a study of slow conductances in pyloric neurons, where variations in the voltage of half-activation of slow calcium and potassium channels were independently shown to replicate phase-maintaining behavior in model neurons. This work strongly suggests that slowly conducting currents are intrinsic cellular mechanisms for delaying neuronal responses, with more emphasis on the dynamics of synaptic activation than a specific ion species.

In a dynamical system, bifurcations can predict the dependence of temporal characteristics of oscillatory regimes near the bifurcation. Inverse-square-root laws have been described for tonic spiking and bursting regimes [49, 56, 69-71]. In type I neuronal dynamics, a saddle-node bifurcation on invariant circle (SNIC) describes a transition from tonic spiking into silence; the period of spiking grows proportionally to $\frac{1}{\sqrt{|\alpha - \alpha^*|}}$ where $\alpha^*$ is the bifurcation value of $\alpha$ and $\alpha > \alpha^*$ [70]. The blue-sky catastrophe, a special case of the saddle-node bifurcation for orbits, imposes an inverse-square-root law on burst duration [49]. This law has previously been shown in bursting neurons approaching the transition to spiking [56]. A saddle-node bifurcation for periodic orbits can control the boundary of bursting activity [49, 56, 71].

## 4.4    Comparison of findings to computational models

Experiments with similar motifs lend further support that network dynamics are determined both by the intrinsic properties of the constituent neurons and the strengths and placement of synapses, although one may extract more influence than the other depending on the particular model. In a symmetrical, mutually reciprocal network of three neurons also derived from the pyloric network central pattern generator, triphasic bursting was demonstrated with three bursting neurons and with one tonic spiking follower and two bursting neurons [23]. Varying the balance of synaptic and cellular influence yielded both bistable and monostable triphasic patterns. While our circuitry differs in that there are no recurrent synapses to the driver, we show how a combination of intrinsic excitability, cellular mechanisms, and asymmetric inhibitory synaptic coupling can generate a stable triphasic pattern with two tonic spiking followers. We confirmed that the bursting order is largely determined by asymmetric coupling in our network model, as endogenously active neurons and silent neurons inhibited into a burst compete to fire first after inhibition.

In modeling experiments with single pyloric neurons, the conductances of the hyperpolarization-activated and potassium currents were not found to be correlated in highly phase-maintaining populations [28]. This effect is in stark contrast with MacLean et al.'s results that the two currents make compensatory adjustments to maintain a constant output profile [29]. However, Soofi et al. suggest that the discrepancy between their findings with those of MacLean may arise from the relative inactivation of the hyperpolarization-activated current in their model. Their results implicate pairs of fast currents in intraburst spike phase maintenance, which may require different dynamics than burst phase maintenance as explored here in a pyloric network motif.

## 4.5   Variability of neuronal properties in circuits

We have shown how the mechanisms may function with follower neurons whose parameters are in the range of endogenous bursting, spiking, and silence. It is feasible, and in the case of the pyloric network it is known, that follower neurons have different channel kinetics from the driver neuron, or even from one another. In the pyloric network, each of the six classes of neurons has its own kinetics and voltage-dependency [72, 73]. In isolation, the follower neurons exhibit qualitatively different dynamics, yet activity is regulated and even stabilized by periodic input from the driver [74]. This functional difference is not to suggest that all synapses in a central pattern generator enforce a pattern or are even requisite for it. For example, removal of the LP synapse to PY has no significant effect on the delay to firing of the PY in response to inhibitory current [54]. This lack of impact of the LP to PY synapse was replicated in Ensembles 3 and 4, where the PY neuron's latency to firing was determined through intrinsic means and had no reliance on inhibition from the LP neuron to preserve its temporal qualities.

Variation occurs not just within particular CPG circuits, but also between animals. The period of activity, density of ion channels in identified neurons, and diameter and length of neurons can vary dramatically between different members of the same species [31-33, 75]. Despite this widespread variation, the phase relationships of active neurons and waveform of the network pattern is essentially consistent across specimens. Our family of mechanisms indicates how identical and temporally scaled output is possible in ensembles where the endogenous activities of the follower neurons are disparate. This evidence supports the ability of specimens with various neuronal properties to yield a consistent, stereotyped pattern regardless of the discrepancies between their cellular dynamics. It also suggests how changes can be orchestrated between neurons in a circuit to maintain functional output robustness when cellular dynamics shift due to modulation or development.

## 4.6   Driver and follower ensembles

The pacemaker neuron (or ensemble) plays a unique role in central pattern generators. It behaves as a clock: the activity of the inhibited follower neurons is constrained to the interburst interval of the inhibiting pacemaking driver, which is determined by pacemaker period and duty cycle. This property allows the cellular mechanisms for control of temporal characteristics to be generalized to significantly larger networks than presented here. In our model, the activity of the two follower neurons evenly divides the interburst interval of the driver. Three follower neurons could evenly divide the interburst interval if coordinated with lower duty cycles, as could any number of followers up to some maximum. This upper bound would be determined by the minimum delay to bursting and duration of the bursts, which is typically a characteristic of cellular slow dynamics and time constants. Of course, Mechanism 1 could be applied to the driver to increase its interburst interval such that a given number of bursts of a given duration will fit. A circuit composed of a driver neuron with reciprocally inhibitory follower neurons is not unique to crustaceans and provides much insight to circuit dynamics that transcends any particular species or function [46]. The lamprey locomotory CPG preserves phase with bilateral synapses [2, 76], and the leech locomotory CPG is similarly structured with the addition of recurrent inhibitory synapses to all neurons [77]. This fundamental motif is also applied to models of recurrent cortical circuits [78], which share many dynamical and structural properties with motor CPGs [79].

Small circuits structured according to this motif may interface to form a network of networks with flexible oscillatory output in a range of timescales. In crayfish swimming, for example, individual networks that control swimmerets must be coordinated to produce a metachronal wave [5, 80]. The stomatogastric system possesses extensive interconnections between the gastric and pyloric CPGs [81-84]. Cortical activity also arises from the interactions of many small networks characterized by a pacemaker with reciprocally inhibitory followers [85]. Our mechanisms for phase maintenance in a single motif provide a possible foundation for understanding the temporal characteristics of these networks of networks.

# 6    Conflict of Interest

The authors declare that the research was conducted in the absence of any commercial or financial relationships that could be construed as a potential conflict of interest.

# 7    Author Contributions

GO, WB, and GC conceived the project. GC oversaw the project scope and project conception. GO, AW, WB, and GC designed models and algorithms, conducted numerical experiments and data analysis, and prepared the manuscript. WB provided support for the development of model algorithms, project design and data analysis, and manuscript preparation. DK conducted numerical experiments and data analysis. All authors read and approved the final manuscript.

## 8    Funding

This work was supported by NIH grants R01NS115209 and R21NS111355 and Brains and Behavior program of Georgia State University to GC. The funders had no role in study design, data collection and analysis, decision to publish, or preparation of the manuscript.

**Data Availability Statement**

The datasets generated for this study are available on request to the corresponding author. The MATLAB code used to produce presented results will be available ModelDB repository and on the website of the corresponding author upon acceptance of the manuscript for publication.

Supplementary Table 1. Used parameters for the studies shown in Figures 2-7 listed by ensemble. A separate red-to-green background color-code was used based on the relative magnitude of each of the following: Ensemble, Trajectory Number, Cycle Period, $\theta_h$, & $\theta_{K2}$.

| Ensemble | Traj. # | Period (s) | $\theta_h$ (V) AB | $\theta_h$ (V) LP | $\theta_h$ (V) PY | $\theta_{K2}$ (V) AB | $\theta_{K2}$ (V) LP | $\theta_{K2}$ (V) PY |
|---|---|---|---|---|---|---|---|---|
| 1 | 1 | 0.500 | 0.041230000000 | 0.041230000000 | 0.041230000000 | -0.0041000 | -0.0041000 | -0.0041000 |
| 1 | 2 | 0.600 | 0.041270000000 | 0.041270000000 | 0.041270000000 | -0.0050000 | -0.0050000 | -0.0050000 |
| 1 | 3 | 0.767 | 0.041307000000 | 0.041307000000 | 0.041307000000 | -0.0065000 | -0.0065000 | -0.0065000 |
| 1 | 4 | 1.167 | 0.041331650000 | 0.041331650000 | 0.041331650000 | -0.0080000 | -0.0080000 | -0.0080000 |
| 1 | 5 | 1.967 | 0.041345950000 | 0.041345950000 | 0.041345950000 | -0.0093000 | -0.0093000 | -0.0093000 |
| 1 | 6 | 2.467 | 0.041349165000 | 0.041349165000 | 0.041349165000 | -0.0096400 | -0.0096400 | -0.0096400 |
| 1 | 7 | 4.433 | 0.041353807000 | 0.041353807000 | 0.041353807000 | -0.0101700 | -0.0101700 | -0.0101700 |
| 1 | 8 | 6.400 | 0.041355101290 | 0.041355101290 | 0.041355101290 | -0.0103267 | -0.0103267 | -0.0103267 |
| 1 | 9 | 8.433 | 0.041355654800 | 0.041355654800 | 0.041355654800 | -0.0103950 | -0.0103950 | -0.0103950 |
| 1 | 10 | 10.367 | 0.041355935900 | 0.041355935900 | 0.041355935900 | -0.0104300 | -0.0104300 | -0.0104300 |
| 1 | 11 | 12.267 | 0.041356095800 | 0.041356095800 | 0.041356095800 | -0.0104500 | -0.0104500 | -0.0104500 |
| 1 | 12 | 14.400 | 0.041356210580 | 0.041356210580 | 0.041356210580 | -0.0104644 | -0.0104644 | -0.0104644 |
| 1 | 13 | 16.233 | 0.041356278990 | 0.041356278990 | 0.041356278990 | -0.0104730 | -0.0104730 | -0.0104730 |
| 1 | 14 | 18.300 | 0.041356329840 | 0.041356329840 | 0.041356329840 | -0.0104794 | -0.0104794 | -0.0104794 |
| 1 | 15 | 20.200 | 0.041356366350 | 0.041356366350 | 0.041356366350 | -0.0104840 | -0.0104840 | -0.0104840 |
| 1 | 16 | 22.200 | 0.041356394113 | 0.041356394113 | 0.041356394113 | -0.0104875 | -0.0104875 | -0.0104875 |
| 1 | 17 | 24.200 | 0.041356415519 | 0.041356415519 | 0.041356415519 | -0.0104902 | -0.0104902 | -0.0104902 |
| 1 | 18 | 26.167 | 0.041356432950 | 0.041356432950 | 0.041356432950 | -0.0104924 | -0.0104924 | -0.0104924 |
| 1 | 19 | 28.067 | 0.041356445626 | 0.041356445626 | 0.041356445626 | -0.0104940 | -0.0104940 | -0.0104940 |
| 1 | 20 | 30.000 | 0.041356461450 | 0.041356461450 | 0.041356461450 | -0.0104960 | -0.0104960 | -0.0104960 |
| 2 | 1 | 0.500 | 0.041230000000 | 0.041500000000 | 0.041500000000 | -0.0041000 | -0.0041000 | -0.0041000 |
| 2 | 2 | 0.600 | 0.041270000000 | 0.041500000000 | 0.041500000000 | -0.0050000 | -0.0050000 | -0.0050000 |
| 2 | 3 | 0.767 | 0.041307000000 | 0.041500000000 | 0.041500000000 | -0.0065000 | -0.0065000 | -0.0065000 |
| 2 | 4 | 1.167 | 0.041331650000 | 0.041500000000 | 0.041500000000 | -0.0080000 | -0.0080000 | -0.0080000 |
| 2 | 5 | 1.967 | 0.041345950000 | 0.041500000000 | 0.041500000000 | -0.0093000 | -0.0093000 | -0.0093000 |
| 2 | 6 | 2.467 | 0.041349165000 | 0.041500000000 | 0.041500000000 | -0.0096400 | -0.0096400 | -0.0096400 |
| 2 | 7 | 4.433 | 0.041353807000 | 0.041500000000 | 0.041500000000 | -0.0101700 | -0.0101700 | -0.0101700 |
| 2 | 8 | 6.400 | 0.041355101290 | 0.041500000000 | 0.041500000000 | -0.0103267 | -0.0103267 | -0.0103267 |
| 2 | 9 | 8.433 | 0.041355654800 | 0.041500000000 | 0.041500000000 | -0.0103950 | -0.0103950 | -0.0103950 |
| 2 | 10 | 10.367 | 0.041355935900 | 0.041500000000 | 0.041500000000 | -0.0104300 | -0.0104300 | -0.0104300 |
| 2 | 11 | 12.267 | 0.041356095800 | 0.041500000000 | 0.041500000000 | -0.0104500 | -0.0104500 | -0.0104500 |
| 2 | 12 | 14.400 | 0.041356210580 | 0.041500000000 | 0.041500000000 | -0.0104644 | -0.0104644 | -0.0104644 |
| 2 | 13 | 16.233 | 0.041356278990 | 0.041500000000 | 0.041500000000 | -0.0104730 | -0.0104730 | -0.0104730 |
| 2 | 14 | 18.300 | 0.041356329840 | 0.041500000000 | 0.041500000000 | -0.0104794 | -0.0104794 | -0.0104794 |
| 2 | 15 | 20.200 | 0.041356366350 | 0.041500000000 | 0.041500000000 | -0.0104840 | -0.0104840 | -0.0104840 |
| 2 | 16 | 22.200 | 0.041356394113 | 0.041500000000 | 0.041500000000 | -0.0104875 | -0.0104875 | -0.0104875 |
| 2 | 17 | 24.200 | 0.041356415519 | 0.041500000000 | 0.041500000000 | -0.0104902 | -0.0104902 | -0.0104902 |
| 2 | 18 | 26.167 | 0.041356432950 | 0.041500000000 | 0.041500000000 | -0.0104924 | -0.0104924 | -0.0104924 |
| 2 | 19 | 28.067 | 0.041356445626 | 0.041500000000 | 0.041500000000 | -0.0104940 | -0.0104940 | -0.0104940 |
| 2 | 20 | 30.000 | 0.041356461450 | 0.041500000000 | 0.041500000000 | -0.0104960 | -0.0104960 | -0.0104960 |
| 3 | 1 | 0.500 | 0.041230000000 | 0.040000000000 | 0.041340000000 | -0.0041000 | -0.0107000 | -0.0106999 |
| 3 | 2 | 0.600 | 0.041270000000 | 0.040000000000 | 0.041349000000 | -0.0050000 | -0.0107000 | -0.0106999 |
| 3 | 3 | 0.767 | 0.041307000000 | 0.040000000000 | 0.041355000000 | -0.0065000 | -0.0107000 | -0.0106999 |
| 3 | 4 | 1.167 | 0.041331650000 | 0.040000000000 | 0.041357500000 | -0.0080000 | -0.0107000 | -0.0106999 |
| 3 | 5 | 1.985 | 0.041345950000 | 0.040000000000 | 0.041357930000 | -0.0093000 | -0.0107000 | -0.0106999 |
| 3 | 6 | 2.467 | 0.041349165000 | 0.040000000000 | 0.041357977000 | -0.0096400 | -0.0107000 | -0.0106999 |
| 3 | 7 | 4.433 | 0.041353807000 | 0.040000000000 | 0.041358025000 | -0.0101700 | -0.0107000 | -0.0106999 |
| 3 | 8 | 6.400 | 0.041355101290 | 0.040000000000 | 0.041358035000 | -0.0103267 | -0.0107000 | -0.0106999 |
| 3 | 9 | 8.600 | 0.041355695000 | 0.040000000000 | 0.041358040000 | -0.0104000 | -0.0107000 | -0.0106999 |
| 3 | 10 | 10.367 | 0.041355935900 | 0.040000000000 | 0.041358042500 | -0.0104300 | -0.0107000 | -0.0106999 |
| 3 | 11 | 12.267 | 0.041356095800 | 0.040000000000 | 0.041358043600 | -0.0104500 | -0.0107000 | -0.0106999 |
| 3 | 12 | 14.400 | 0.041356210580 | 0.040000000000 | 0.041358044400 | -0.0104644 | -0.0107000 | -0.0106999 |
| 3 | 13 | 16.233 | 0.041356278990 | 0.040000000000 | 0.041358044800 | -0.0104730 | -0.0107000 | -0.0106999 |
| 3 | 14 | 18.300 | 0.041356329840 | 0.040000000000 | 0.041358045200 | -0.0104794 | -0.0107000 | -0.0106999 |
| 3 | 15 | 20.200 | 0.041356366350 | 0.040000000000 | 0.041358045400 | -0.0104840 | -0.0107000 | -0.0106999 |
| 3 | 16 | 22.200 | 0.041356394113 | 0.040000000000 | 0.041358045600 | -0.0104875 | -0.0107000 | -0.0106999 |
| 3 | 17 | 24.200 | 0.041356415519 | 0.040000000000 | 0.041358045800 | -0.0104902 | -0.0107000 | -0.0106999 |
| 3 | 18 | 26.167 | 0.041356432950 | 0.040000000000 | 0.041358045900 | -0.0104924 | -0.0107000 | -0.0106999 |
| 3 | 19 | 28.067 | 0.041356445626 | 0.040000000000 | 0.041358046000 | -0.0104940 | -0.0107000 | -0.0106999 |
| 3 | 20 | 30.000 | 0.041356461450 | 0.040000000000 | 0.041358046050 | -0.0104960 | -0.0107000 | -0.0106999 |
| 4 | 1 | 0.500 | 0.041230000000 | 0.041500000000 | 0.041340000000 | -0.0041000 | -0.0041000 | -0.0106999 |
| 4 | 2 | 0.600 | 0.041270000000 | 0.041500000000 | 0.041349000000 | -0.0050000 | -0.0050000 | -0.0106999 |
| 4 | 3 | 0.767 | 0.041307000000 | 0.041500000000 | 0.041355000000 | -0.0065000 | -0.0065000 | -0.0106999 |
| 4 | 4 | 1.167 | 0.041331650000 | 0.041500000000 | 0.041357500000 | -0.0080000 | -0.0080000 | -0.0106999 |
| 4 | 5 | 1.967 | 0.041345950000 | 0.041500000000 | 0.041357930000 | -0.0093000 | -0.0093000 | -0.0106999 |
| 4 | 6 | 2.467 | 0.041349165000 | 0.041500000000 | 0.041357977000 | -0.0096400 | -0.0096400 | -0.0106999 |
| 4 | 7 | 4.433 | 0.041353807000 | 0.041500000000 | 0.041358025000 | -0.0101700 | -0.0101700 | -0.0106999 |
| 4 | 8 | 6.400 | 0.041355101290 | 0.041500000000 | 0.041358035000 | -0.0103267 | -0.0103267 | -0.0106999 |
| 4 | 9 | 8.600 | 0.041355695000 | 0.041500000000 | 0.041358040000 | -0.0104000 | -0.0104000 | -0.0106999 |
| 4 | 10 | 10.367 | 0.041355935900 | 0.041500000000 | 0.041358042500 | -0.0104300 | -0.0104300 | -0.0106999 |
| 4 | 11 | 12.267 | 0.041356095800 | 0.041500000000 | 0.041358043600 | -0.0104500 | -0.0104500 | -0.0106999 |
| 4 | 12 | 14.400 | 0.041356210580 | 0.041500000000 | 0.041358044400 | -0.0104644 | -0.0104644 | -0.0106999 |
| 4 | 13 | 16.233 | 0.041356278990 | 0.041500000000 | 0.041358044800 | -0.0104730 | -0.0104730 | -0.0106999 |
| 4 | 14 | 18.300 | 0.041356329840 | 0.041500000000 | 0.041358045200 | -0.0104794 | -0.0104794 | -0.0106999 |
| 4 | 15 | 20.200 | 0.041356366350 | 0.041500000000 | 0.041358045400 | -0.0104840 | -0.0104840 | -0.0106999 |
| 4 | 16 | 22.200 | 0.041356394113 | 0.041500000000 | 0.041358045600 | -0.0104875 | -0.0104875 | -0.0106999 |
| 4 | 17 | 24.200 | 0.041356415519 | 0.041500000000 | 0.041358045800 | -0.0104902 | -0.0104920 | -0.0106999 |
| 4 | 18 | 26.167 | 0.041356432950 | 0.041500000000 | 0.041358045900 | -0.0104924 | -0.0104924 | -0.0106999 |
| 4 | 19 | 28.067 | 0.041356445626 | 0.041500000000 | 0.041358046000 | -0.0104940 | -0.0104940 | -0.0106999 |
| 4 | 20 | 30.000 | 0.041356461450 | 0.041500000000 | 0.041358046050 | -0.0104960 | -0.0104960 | -0.0106999 |

**Supplementary Table 2.** Used parameters for the studies shown in Figure 8 listed by panel. A separate red-to-green background color-code was used based on the relative magnitude of each of the following: Ensemble, Trajectory Number, Cycle Period, $\theta_h$, & $\theta_{K2}$.

| Panel | Ensemble | Traj. # | Period (s) | $\theta_h$ (V) | | | $\theta_{K2}$ (V) | | |
|---|---|---|---|---|---|---|---|---|---|
| | | | | AB | LP | PY | AB | LP | PY |
| A | 1 | 1 | 0.500 | 0.041230000000 | 0.041356461450 | 0.041356461450 | -0.0041000 | -0.0041000 | -0.0104960 |
| A | 1 | 2 | 0.600 | 0.041270000000 | 0.041356461450 | 0.041356461450 | -0.0050000 | -0.0050000 | -0.0104960 |
| A | 1 | 3 | 0.767 | 0.041307000000 | 0.041356461450 | 0.041356461450 | -0.0065000 | -0.0065000 | -0.0104960 |
| A | 1 | 4 | 1.167 | 0.041331650000 | 0.041356461450 | 0.041356461450 | -0.0080000 | -0.0080000 | -0.0104960 |
| A | 1 | 5 | 1.967 | 0.041345950000 | 0.041356461450 | 0.041356461450 | -0.0093000 | -0.0093000 | -0.0104960 |
| A | 1 | 6 | 2.467 | 0.041349165000 | 0.041356461450 | 0.041356461450 | -0.0096400 | -0.0096400 | -0.0104960 |
| A | 1 | 7 | 4.433 | 0.041353807000 | 0.041356461450 | 0.041356461450 | -0.0101700 | -0.0101700 | -0.0104960 |
| A | 1 | 8 | 6.400 | 0.041355101290 | 0.041356461450 | 0.041356461450 | -0.0103267 | -0.0103267 | -0.0104960 |
| A | 1 | 9 | 8.433 | 0.041355654800 | 0.041356461450 | 0.041356461450 | -0.0103950 | -0.0103950 | -0.0104960 |
| A | 1 | 10 | 10.367 | 0.041355933900 | 0.041356461450 | 0.041356461450 | -0.0104300 | -0.0104300 | -0.0104960 |
| A | 1 | 11 | 12.267 | 0.041356095800 | 0.041356461450 | 0.041356461450 | -0.0104500 | -0.0104500 | -0.0104960 |
| A | 1 | 12 | 14.400 | 0.041356210580 | 0.041356461450 | 0.041356461450 | -0.0104644 | -0.0104644 | -0.0104960 |
| A | 1 | 13 | 16.233 | 0.041356278990 | 0.041356461450 | 0.041356461450 | -0.0104730 | -0.0104730 | -0.0104960 |
| A | 1 | 14 | 18.300 | 0.041356329840 | 0.041356461450 | 0.041356461450 | -0.0104794 | -0.0104794 | -0.0104960 |
| A | 1 | 15 | 20.200 | 0.041356363500 | 0.041356461450 | 0.041356461450 | -0.0104840 | -0.0104840 | -0.0104960 |
| A | 1 | 16 | 22.200 | 0.041356394113 | 0.041356461450 | 0.041356461450 | -0.0104875 | -0.0104875 | -0.0104960 |
| A | 1 | 17 | 24.200 | 0.041356415519 | 0.041356461450 | 0.041356461450 | -0.0104902 | -0.0104902 | -0.0104960 |
| A | 1 | 18 | 26.167 | 0.041356432950 | 0.041356461450 | 0.041356461450 | -0.0104924 | -0.0104924 | -0.0104960 |
| A | 1 | 19 | 28.067 | 0.041356445626 | 0.041356461450 | 0.041356461450 | -0.0104940 | -0.0104940 | -0.0104960 |
| A | 1 | 20 | 30.000 | 0.041356461450 | 0.041356461450 | 0.041356461450 | -0.0104960 | -0.0104960 | -0.0104960 |
| B | 3 | 1 | 0.500 | 0.041230000000 | 0.040000000000 | 0.041340000000 | -0.0041000 | -0.0107000 | -0.0106999 |
| B | 3 | 2 | 0.600 | 0.041270000000 | 0.040000000000 | 0.041349000000 | -0.0050000 | -0.0107000 | -0.0106999 |
| B | 3 | 3 | 0.767 | 0.041307000000 | 0.040000000000 | 0.041355000000 | -0.0065000 | -0.0107000 | -0.0106999 |
| B | 3 | 4 | 1.167 | 0.041331650000 | 0.040000000000 | 0.041357500000 | -0.0080000 | -0.0107000 | -0.0106999 |
| B | 3 | 5 | 1.967 | 0.041345950000 | 0.040000000000 | 0.041357930000 | -0.0093000 | -0.0107000 | -0.0106999 |
| B | 3 | 6 | 2.467 | 0.041349165000 | 0.040000000000 | 0.041357977000 | -0.0096400 | -0.0107000 | -0.0106999 |
| B | 3 | 7 | 4.433 | 0.041353807000 | 0.040000000000 | 0.041358025000 | -0.0101700 | -0.0107000 | -0.0106999 |
| B | 3 | 8 | 6.400 | 0.041355101290 | 0.040000000000 | 0.041358037000 | -0.0103267 | -0.0107000 | -0.0106999 |
| B | 3 | 9 | 8.433 | 0.041355569000 | 0.040000000000 | 0.041358040000 | -0.0104000 | -0.0107000 | -0.0106999 |
| B | 3 | 10 | 10.367 | 0.041355933900 | 0.040000000000 | 0.041358042500 | -0.0104300 | -0.0107000 | -0.0106999 |
| B | 3 | 11 | 12.267 | 0.041356095800 | 0.040000000000 | 0.041358043600 | -0.0104500 | -0.0107000 | -0.0106999 |
| B | 3 | 12 | 14.400 | 0.041356210580 | 0.040000000000 | 0.041358044400 | -0.0104644 | -0.0107000 | -0.0106999 |
| B | 3 | 13 | 16.233 | 0.041356278990 | 0.040000000000 | 0.041358045200 | -0.0104730 | -0.0107000 | -0.0106999 |
| B | 3 | 14 | 18.300 | 0.041356329840 | 0.040000000000 | 0.041358045400 | -0.0104794 | -0.0107000 | -0.0106999 |
| B | 3 | 15 | 20.200 | 0.041356363500 | 0.040000000000 | 0.041358045400 | -0.0104840 | -0.0107000 | -0.0106999 |
| B | 3 | 16 | 22.200 | 0.041356394113 | 0.040000000000 | 0.041358045600 | -0.0104875 | -0.0107000 | -0.0106999 |
| B | 3 | 17 | 24.200 | 0.041356415519 | 0.040000000000 | 0.041358045900 | -0.0104902 | -0.0107000 | -0.0106999 |
| B | 3 | 18 | 26.167 | 0.041356432950 | 0.040000000000 | 0.041358045900 | -0.0104924 | -0.0107000 | -0.0106999 |
| B | 3 | 19 | 28.067 | 0.041356445626 | 0.040000000000 | 0.041358046000 | -0.0104940 | -0.0107000 | -0.0106999 |
| B | 3 | 20 | 30.000 | 0.041356461450 | 0.040000000000 | 0.041358046050 | -0.0104960 | -0.0107000 | -0.0106999 |
| C | 4 | 1 | 0.500 | 0.041223000000 | 0.041500000000 | 0.041358046050 | -0.0041000 | -0.0041000 | -0.0106999 |
| C | 4 | 2 | 0.600 | 0.041270000000 | 0.041500000000 | 0.041358046050 | -0.0050000 | -0.0050000 | -0.0106999 |
| C | 4 | 3 | 0.767 | 0.041307000000 | 0.041500000000 | 0.041358046050 | -0.0065000 | -0.0065000 | -0.0106999 |
| C | 4 | 4 | 1.167 | 0.041331650000 | 0.041500000000 | 0.041358046050 | -0.0080000 | -0.0080000 | -0.0106999 |
| C | 4 | 5 | 1.967 | 0.041345950000 | 0.041500000000 | 0.041358046050 | -0.0093000 | -0.0093000 | -0.0106999 |
| C | 4 | 6 | 2.467 | 0.041349165000 | 0.041500000000 | 0.041358046050 | -0.0096400 | -0.0096400 | -0.0106999 |
| C | 4 | 7 | 4.433 | 0.041353807000 | 0.041500000000 | 0.041358046050 | -0.0101700 | -0.0101700 | -0.0106999 |
| C | 4 | 8 | 6.400 | 0.041355101290 | 0.041500000000 | 0.041358046050 | -0.0103267 | -0.0103267 | -0.0106999 |
| C | 4 | 9 | 8.433 | 0.041355569000 | 0.041500000000 | 0.041358046050 | -0.0104000 | -0.0104000 | -0.0106999 |
| C | 4 | 10 | 10.367 | 0.041355933900 | 0.041500000000 | 0.041358046050 | -0.0104300 | -0.0104300 | -0.0106999 |
| C | 4 | 11 | 12.267 | 0.041356095800 | 0.041500000000 | 0.041358046050 | -0.0104500 | -0.0104500 | -0.0106999 |
| C | 4 | 12 | 14.400 | 0.041356210580 | 0.041500000000 | 0.041358046050 | -0.0104644 | -0.0104644 | -0.0106999 |
| C | 4 | 13 | 16.233 | 0.041356278990 | 0.041500000000 | 0.041358046050 | -0.0104730 | -0.0104730 | -0.0106999 |
| C | 4 | 14 | 18.300 | 0.041356329840 | 0.041500000000 | 0.041358046050 | -0.0104794 | -0.0104794 | -0.0106999 |
| C | 4 | 15 | 20.200 | 0.041356363500 | 0.041500000000 | 0.041358046050 | -0.0104840 | -0.0104840 | -0.0106999 |
| C | 4 | 16 | 22.200 | 0.041356394113 | 0.041500000000 | 0.041358046050 | -0.0104875 | -0.0104875 | -0.0106999 |
| C | 4 | 17 | 24.200 | 0.041356415519 | 0.041500000000 | 0.041358046050 | -0.0104902 | -0.0104920 | -0.0106999 |
| C | 4 | 18 | 26.167 | 0.041356432950 | 0.041500000000 | 0.041358046050 | -0.0104924 | -0.0104924 | -0.0106999 |
| C | 4 | 19 | 28.067 | 0.041356445626 | 0.041500000000 | 0.041358046050 | -0.0104940 | -0.0104940 | -0.0106999 |
| C | 4 | 20 | 30.000 | 0.041356461450 | 0.041500000000 | 0.041358046050 | -0.0104960 | -0.0104960 | -0.0106999 |
| D | 4 | 1 | 0.500 | 0.041223000000 | 0.041500000000 | 0.041340000000 | -0.0041000 | -0.0104960 | -0.0106999 |
| D | 4 | 2 | 0.600 | 0.041270000000 | 0.041500000000 | 0.041349000000 | -0.0050000 | -0.0104960 | -0.0106999 |
| D | 4 | 3 | 0.767 | 0.041307000000 | 0.041500000000 | 0.041355000000 | -0.0065000 | -0.0104960 | -0.0106999 |
| D | 4 | 4 | 1.167 | 0.041331650000 | 0.041500000000 | 0.041357500000 | -0.0080000 | -0.0104960 | -0.0106999 |
| D | 4 | 5 | 1.967 | 0.041345950000 | 0.041500000000 | 0.041357930000 | -0.0093000 | -0.0104960 | -0.0106999 |
| D | 4 | 6 | 2.467 | 0.041349165000 | 0.041500000000 | 0.041357977000 | -0.0096400 | -0.0104960 | -0.0106999 |
| D | 4 | 7 | 4.433 | 0.041353807000 | 0.041500000000 | 0.041358025000 | -0.0101700 | -0.0104960 | -0.0106999 |
| D | 4 | 8 | 6.400 | 0.041355101290 | 0.041500000000 | 0.041358035000 | -0.0103267 | -0.0104960 | -0.0106999 |
| D | 4 | 9 | 8.433 | 0.041355569000 | 0.041500000000 | 0.041358040000 | -0.0104000 | -0.0104960 | -0.0106999 |
| D | 4 | 10 | 10.367 | 0.041355933900 | 0.041500000000 | 0.041358042500 | -0.0104300 | -0.0104960 | -0.0106999 |
| D | 4 | 11 | 12.267 | 0.041356095800 | 0.041500000000 | 0.041358043600 | -0.0104500 | -0.0104960 | -0.0106999 |
| D | 4 | 12 | 14.400 | 0.041356210580 | 0.041500000000 | 0.041358044400 | -0.0104644 | -0.0104960 | -0.0106999 |
| D | 4 | 13 | 16.233 | 0.041356278990 | 0.041500000000 | 0.041358045200 | -0.0104730 | -0.0104960 | -0.0106999 |
| D | 4 | 14 | 18.300 | 0.041356329840 | 0.041500000000 | 0.041358045400 | -0.0104794 | -0.0104960 | -0.0106999 |
| D | 4 | 15 | 20.200 | 0.041356363500 | 0.041500000000 | 0.041358045400 | -0.0104840 | -0.0104960 | -0.0106999 |
| D | 4 | 16 | 22.200 | 0.041356394113 | 0.041500000000 | 0.041358045600 | -0.0104875 | -0.0104960 | -0.0106999 |
| D | 4 | 17 | 24.200 | 0.041356415519 | 0.041500000000 | 0.041358045900 | -0.0104902 | -0.0104960 | -0.0106999 |
| D | 4 | 18 | 26.167 | 0.041356432950 | 0.041500000000 | 0.041358045900 | -0.0104924 | -0.0104960 | -0.0106999 |
| D | 4 | 19 | 28.067 | 0.041356445626 | 0.041500000000 | 0.041358046000 | -0.0104940 | -0.0104960 | -0.0106999 |
| D | 4 | 20 | 30.000 | 0.041356461450 | 0.041500000000 | 0.041358046050 | -0.0104960 | -0.0104960 | -0.0106999 |